\documentclass[12pt]{article}%
\usepackage{adjustbox}
\usepackage{bm}
\usepackage{amsfonts}
\usepackage[colorlinks,citecolor=blue,urlcolor=blue,hyperfootnotes=false]%
{hyperref}

\usepackage{amssymb}
\usepackage{xcolor}
 \usepackage{xr}
\usepackage{floatrow}
\usepackage{apptools}
\usepackage[bottom]{footmisc}
\usepackage[hyperref]{ntheorem}
\usepackage{enumitem}
\usepackage{natbib}
\usepackage{graphicx}
\usepackage[doublespacing,nodisplayskipstretch ]{setspace}
\usepackage[margin=1in, a4paper]{geometry}
\usepackage{booktabs}
\usepackage{threeparttable}
\usepackage{amsmath}
\usepackage{lscape}
\usepackage{scrextend}
\usepackage{setspace}
\usepackage{relsize}
\usepackage{multicol}
\usepackage{chngcntr}
\usepackage{etoolbox}
\usepackage{tabularx}
\usepackage[english]{babel}
\usepackage{multirow}
\usepackage{ntheorem}
\usepackage{booktabs}
\usepackage[labelfont=bf]{caption}
\usepackage{float}
\usepackage{titlesec}
\usepackage{array}
\usepackage{tikz}
\usepackage{color}
\usepackage{afterpage}
\usepackage{framed}
\usepackage{subcaption}
\usepackage[titletoc,title]{appendix}
\usepackage[section]{placeins}
\providecommand{\U}[1]{\protect\rule{.1in}{.1in}}
\allowdisplaybreaks

\theoremstyle{plain}
\newtheorem{theorem}{Theorem}

\newtheorem{lemma}[theorem]{Lemma}
\newtheorem{definition}{Definition}
\newtheorem{assumption}{Assumption}
\newtheorem{example}{Example}
\newtheorem{remark}{Remark}
\newtheorem{algorithm}{Algorithm}

\AtAppendix{\counterwithin{theorem}{section}}
\AtAppendix{\counterwithin{assumption}{section}}

\deffootnote{1em}{1.6em}{\thefootnotemark \enskip}

\newcommand\independent{\protect\mathpalette{\protect\independentT}{\perp}}
\def\independentT#1#2{\mathrel{\rlap{$#1#2$}\mkern2mu{#1#2}}}

\definecolor{lightgray}{gray}{0.9}
\AtBeginDocument{
	\addtolength\abovedisplayskip{-0.4\baselineskip}
}
\def\ubar#1{\underline{#1}}
\DeclareMathOperator{\R}{\mathbb{R}}
\DeclareMathOperator{\E}{\mathbb{E}}

\DeclareMathOperator{\D}{\mathcal{D}}
\DeclareMathOperator{\LL}{\mathcal{L}}
\DeclareMathOperator{\F}{\mathcal{F}}

\DeclareMathOperator{\Z}{\mathcal{Z}}
\DeclareMathOperator{\Y}{\mathcal{Y}}
\DeclareMathOperator{\X}{\mathcal{X}}
\DeclareMathOperator{\W}{\mathcal{W}}

\DeclareMathOperator{\J}{\mathcal{J}}
\DeclareMathOperator{\Pp}{\mathcal{P}}
\DeclareMathOperator{\PP}{\mathbb{P}}

\newcommand{\norm}[1]{\left\lVert#1\right\rVert}

\DeclareMathOperator\supp{Supp}

\def\independentT#1#2{\mathrel{\rlap{$#1#2$}\mkern2mu{#1#2}}}

\allowdisplaybreaks

\begin{document}

\title{Two Sample Unconditional Quantile Effect}	
\author{Atsushi Inoue \thanks{Department of Economics, Vanderbilt University. Email: atsushi.inoue@vanderbilt.edu}\\  Vanderbilt University \and  Tong Li  \thanks{Department of Economics, Vanderbilt University. Email: tong.li@vanderbilt.edu} \\Vanderbilt University \and Qi Xu \thanks{Department of Economics, Vanderbilt University. Email: qi.xu.1@vanderbilt.edu} \\Vanderbilt University} 
\maketitle

\begin{abstract}
This paper proposes a new framework to evaluate \textit{unconditional quantile effects} (UQE) in a data combination model.  The UQE measures the effect of a marginal counterfactual change in the unconditional distribution of a covariate on quantiles of the unconditional distribution of a target outcome. Under rank similarity and conditional independence assumptions, we provide a set of identification results for UQEs when the target covariate is continuously distributed and when it is discrete, respectively.  Based on these identification results, we propose semiparametric estimators and establish their large sample properties under primitive conditions. Applying our method to a variant of Mincer's earnings function, we study the counterfactual quantile effect of actual work experience on income.
\end{abstract}
\pagebreak

\section{Introduction}
Missing data is a ubiquitous problem in empirical studies. Consider the scenario where a researcher is interested in conducting counterfactual analysis on a target variable, but it is entirely missing from the dataset of interest. In such circumstances, counterfactual policy effects cannot be identified from the primary dataset alone, and therefore, external information and/or stronger identifying assumptions are necessary.  In this paper, we utilize both to achieve identification.  Specifically, we focus on the situation where the missing variable can be found in another dataset and the information from which can be used to recover target policy parameters in the population of interest, under a set of commonly assumed restrictions on both the data structure and the model primitives.  

To fix ideas, consider the following example. Suppose we are interested in studying the effect of a counterfactual change in the distribution of actual labor market experience on some distributional feature of yearly earnings. Our main dataset does not record respondents' work history, and therefore, we cannot recover their actual labor market experience. Suppose the variable is available from a second dataset, but it may not be a reliable source of information on income or it may not be representative of the target population we aim to analyze.  In this case, we would benefit from combining information from both samples to identify and estimate our parameter of interest.


Research on counterfactual policy effects under data combination is scarce. Our paper fills this gap by proposing a new framework that accommodates such a data structure.  In this paper, we focus on one particular type of counterfactual policy effects, the \textit{unconditional quantile effect} (UQE). It measures the effect of a marginal change in the unconditional distribution of a single covariate on the quantiles of a target outcome. We provide identification results for UQE under various types of marginal distributional change.  The key insight of our identification strategy is that some covariates present in both datasets can be excluded from the outcome equation, which would provide a source of exogenous variations that allows us to recover the joint distribution of missing variables, otherwise not identified using the two samples separately. 

The second contribution of the paper is to propose novel semiparametric estimators based on these identification results.  Departing from the literature on the estimation of counterfactual quantile effects\textemdash see, e.g. \cite{firpo2009unconditional}, \cite{sasaki2020unconditional}, etc.\textemdash which focuses primarily on the \textit{marginal location shift} (MLS) of a covariate, we provide estimators of UQE under two general types of counterfactual distributional changes, namely the \textit{marginal distributional shift} (MDS) and the \textit{marginal quantile shift} (MQS),\footnote{The precise definitions of MLS, MDS, and MQS are given in Section \ref{identification}.} the latter of which includes MLS as a special case. To the best of our knowledge, large sample results for these two cases are new to the literature. We apply these results to study a variant of Mincer's earnings function. Using data from Integrated Public Use Microdata Sample (IPUMS) as our main data source and the Panel Study of Income Dynamics (PSID) as the auxiliary sample, we investigate the counterfactual income effect of actual work experience. The effect profiles with MDS and MQS are found to be similar in shape. 

This paper belongs to the growing literature on the  unconditional policy effect.  Since \cite{firpo2009unconditional} introduced the method of \textit{unconditional quantile regressions} (UQR),  the study of unconditional policy effect has gained much attention. In general, this parameter differs from the one identified by the conditional quantile regression \citep{koenker1978regression}, where marginal effects on the conditional quantile are the locus of attention. Applied researcher are often interested in the shifts in the quantiles of unconditional distribution of a target outcome. For instance, one may take an interest in how wage distribution changes in response to marginal increases in some characteristics of the labor force, such as education level and experience. Conditional quantile regression cannot be applied to address this type of questions, whereas UQR suits the goal.  

\citet{rothe2012partial} generalizes the method of \citet{firpo2009unconditional}, and analyzes a variety of counterfactual policy effects. He formalizes the idea of \textit{ceteris paribus} distributional change and provides extensive results for both fixed and marginal policy shifts. Our identification framework is closely related to his treatment of the latter type. Focusing on the special case of quantile effects, we extend his identification results to a data combination setting and provide novel inference theories specifically tailored to the distinct features of combined samples.  For recent development in this literature, see \citet{martinez2020identification}, \citet{martinez2020sensitivity}, and \citet{sasaki2020unconditional}.   For a comprehensive survey on counterfactual distributions and decomposition methods, see \citet{fortin2011decomposition}. 

Our paper also builds on the econometric methods of data combination.  In economics, this strand of literature stems from the \textit{two-sample instrumental variables} (TSIV) model that was first introduced by \citet{klevmarken1982missing}, \citet{angrist1992effect}, \citet{arellano1992female}, and is later extended by  \citet{ridder2007econometrics}, \citet{inoue2010two}, among others. Conceptually,  the semiparametric data combination model we consider here is different from the traditional missing data problem \citep{robins1994estimation}. It is more closely related to the ``verify-out-of-sample'' model in \citet{chen2008semiparametric}, and also to \citet{imbens1994combining}, \citet{fan2014identifying}, \citet{graham2016efficient}, \citet{hirukawa2020yet}, and \citet{buchinsky2021estimation}, to name a few. 



The paper is organized as follows.  In the next section, we describe the model and assumptions on the data structure. In Section \ref{identification} we introduce the parameter of interest, and then present identification results for continuously distributed and discrete target covariates, respectively.  Section \ref{estimation} discusses the estimation strategy and large sample results. We apply the method to study the income effect of real labor market experience in  Section \ref{application}.    Section \ref{conclusion} concludes.

\section{Setup}\label{setup}

The objective of our paper is to analyze the effect of a counterfactual change in the marginal distribution of the covariate of interest, $ X $, on the quantiles of the target outcome, $ Y $, under data combination.   The precise definition of the counterfactual policy effect is provided in Section \ref{identification}.  When $ X $ is exogenous, and all the variables relevant for analysis are observed from a single data source, counterfactual policy effects can be analyzed either directly by applying tools from \citet{firpo2009unconditional} and \citet{rothe2012partial}, or indirectly by recovering the structural function using standard identification results such as \citet{matzkin2003nonparametric} and \citet{matzkin2007nonparametric}. However, when the variables of interest are scattered among several different data sources, we face a fundamental identification problem: The conditional distribution of $ Y $ given $ X $ is not identified from any single sample. In this case, existing methods do not provide an immediate solution. 

Throughout this paper, we consider the scenario where our  $ Y $ and $ X $ are sourced from two different data sets. The outcome is contained in the \textit{principal} or \textit{main} sample, $ \mathcal{S}_{s} = \{ Y_{i}, Z_{i} \}_{i = 1}^{n_{s}}  $, from the \textit{study} population,  $\mathcal{P}_{s} $. The target covariate  is missing completely from $ \mathcal{S}_{s}  $. However, it is observed in the \textit{auxiliary} sample, $ \mathcal{S}_{a} = \{ X_{i},  Z_{i} \}_{i = 1}^{n_{a}} $, from the \textit{auxiliary} population, $ \mathcal{P}_{a}$, which does not contain observations of $ Y $. 

We now formally describe our structural model. 
 We allow variables from two populations to be determined by different mechanisms. For the study population, 
\begin{align} 
Y_{s} &= g_{s}(X_{s}, Z_{1}, \epsilon_{s}), \label{eq.outcome}  \\
X_{s} &= h_{s}(Z, \eta_{s}), \label{eq.reduced} 
\end{align}
where $ Y_{s} \in \mathcal{Y} \subset \R $ is the potential outcome in the study population, $ \epsilon_{s} \in \mathcal{E} \subset \mathbb{R}^{d_{\epsilon}} $ is a vector of unobserved heterogeneity term.  Equation \eqref{eq.outcome}, links the target outcome,  a scalar variable, $ X_{s} \in \mathcal{X} \subset \mathbb{R} $, and a vector of exogenous variables, $ Z_{1}  $.  Here, $ X_{s} $ is the potential covariate of interest in the study population, which is in turn determined by \eqref{eq.reduced}.   We can think of \eqref{eq.reduced} as the reduced form relationship between $ X_{s} $ and $ Z $,  where  $ Z' := (Z_{1}', Z_{2}')' \in \Z:= \Z_{1} \times \Z_{2} \subset \R^{d_{z}} $ includes both the exogenous variables in the outcome equation and a vector of excluded instrument, $ Z_{2} $.   The vector of instrument, $ Z $, is available in both samples, and therefore, it serves to establish a link between two samples.\footnote{The support of $ Z $ is not restricted. We allow the distribution of $ Z $ to be continuous, discrete, or mixed. Nevertheless, to ease notational burden, we focus on the continuous case exclusively in what follows.}  The model in \eqref{eq.outcome} accommodates general nonseparability between covariates and the unobserved heterogeneity. We do not impose any parametric or shape restriction on $ g_{s}$.   

Variables in the auxiliary population are determined by 
\begin{align*} 
Y_{a} &= g_{a}(X_{a}, Z_{1}, \epsilon_{a}), \\
X_{a} &= h_{a}(Z, \eta_{a}), 
\end{align*}
where $ g_{a} $ and $ h_{a} $ are generally different from $ g_{s} $ and $ h_{s} $. 

Let $ R $ denote the sample membership indicator.  That is, $ R_{i} = 1, $ if $ i $-th draw comes from the study population, $ i = 1,...,n :=n_{s}+n_{a} $.   Let $ Y := RY_{s} + (1-R)Y_{a} $ and $ X := RX_{s} + (1-R)X_{a}  $. If  no variable is missing, we are able to observe $ (Y_{s}, Y_{a}, X_{s}, X_{a}) $.  However, in our context, only $ RY $ and  $  (1-R)X $ are observed. We then construct a pseudo-merged sample $ \mathcal{S} $ using the two data sources as $ \mathcal{S} = \{ R_{i}, R_{i}Y_{i}, (1-R_{i})X_{i},  Z_{i}\}_{i= 1}^{n} $. Let $ A := (R, RY, (1-R)X, Z) $ and $ W := (X, Z_{1}) $ collect the observed variables and the covariates in the outcome equation, respectively.  Throughout the paper, we arrange the data in a way such that $ R_{i}=1 $ for $  i=1,...,n_{s} $ and $ R_{i}=0 $ for $ i=n_{s}+ 1,...,n. $  The merged sample may not correspond to any real-world population. We impose the following set of assumptions on the merged sample so it can mimic a random sample from a pseudo population. These assumptions are largely based on Assumption 1 in \citet{graham2016efficient}.  

\begin{assumption}[Data Structure]\label{assmp.data.struct} \ 
	\begin{enumerate}[label=(\alph*)]
		\item  $ \supp(F_{Z|R =1}) \subset \supp (F_{Z|R =0})   $.
		\item  (i) $  n_{s}/(n_{s} + n_{a}) \to Q_{0} $; (ii) $ R $ follows a Bernoulli distribution, with $ \mathbb{E}[R] = Q_{0}  $. 
		\item There is a unique measurable function $ r(\cdot): \Z \mapsto [0,1] $, such that for all $ z \in \Z$,
		\begin{equation*}
		\dfrac{ f_{Z|R}(z | 1)}{f_{Z|R}(z|0) }  =  \dfrac{1-Q_{0}}{Q_{0}} \dfrac{r(z)}{1-r(z)}.
		\end{equation*}
		\item  (i) $Q_{0}  \in (\epsilon_{1}, 1-\epsilon_{1}), $ for some $ \epsilon_{1} \in (0, 1/2)  $; 
		(ii) $ \epsilon_{2} < r(z) < 1- \epsilon_{2} $ for some $ \epsilon_{2} \in (0, 1/2)  $,  and for all $  \  z \in \mathcal{Z}$.
		\item  $   (X_{s} |Z, R  = 1) \stackrel{d}{=} (X_{a}|Z, R = 0)$.		
	\end{enumerate}
\end{assumption}

Assumption \ref{assmp.data.struct}(a) is a support condition on the commonly observed variables. It ensures that we will be able to find,  for all the observations in the study sample,  comparable units in the auxiliary sample,  
Assumption \ref{assmp.data.struct}(b) imposes a pseudo randomization scheme on $ R $, and therefore, allows us to view the merged data as a random sample from the pseudo-merged population.  Let $ \ell(\cdot) $ denote the conditional likelihood ratio of $ Z $ across two population, i.e. $ \ell(z) :=  f_{Z|R}(z | 1)/f_{Z|R}(z|0) $. Assumption \ref{assmp.data.struct}(c) expresses this likelihood ratio as a function of  $ r(\cdot) $, which plays the role of the ``propensity score'' function of $ R $ given $ Z $.  In our context, this is the probability that one observation belongs to the study population conditional on the value that instrumental variables take. 
The first part of Assumption \ref{assmp.data.struct}(d) indicates that $ n_{s}  $ grows at the same order of magnitude as $ n_{a} $. 
The second part of Assumption \ref{assmp.data.struct}(d) ensures that the pseudo-true merged population is not a  degenerate one conditional on all possible values of $ Z $. By Assumption \ref{assmp.data.struct}(b)--(d) and Bayes' Law, we have  $ r(z) = \PP(D = 1| Z = z), $
and thus, $ r(\cdot) $ can be viewed as the propensity score function.   

Assumption \ref{assmp.data.struct}(e) is a rank similarity condition. It requires the conditional distribution of $ X_{s} $ given $ Z $ in the principal population coincide with that of $ X_{a} $ in the auxiliary population. Assumption \ref{assmp.data.struct}(e) is the only cross-population restriction we impose on our data structure, which means the conditional distribution of $ Y $ given $ (X, Z) $, and therefore, the conditional distribution of $ Y $ given $ Z $ and the marginal distribution of $ Z $ are all allowed to differ across $ \Pp_{s} $ and $ \Pp_{a} $.  This assumption is weaker than Assumption 1(ii) of \citet{graham2016efficient}, as we do not impose a rank similarity condition on the outcome, which would imply  $ F_{Y_{s}|ZR = 1}(\cdot | \cdot)  = F_{Y_{a}| Z R = 0}(\cdot | \cdot)$.


\section{Identification}\label{identification}

In this section, we first introduce the definition of UQE. Then, we develop a set of identification results, for the cases when $ X $ is continuously distributed, and when it is discrete, respectively.

\subsection{Parameter of Interest}
Our definition of the unconditional policy effect depends on the notion of a counterfactual experiment, which is formally defined as follows,

\begin{definition}[Counterfactual Experiment] \ Let $ \phi := (\widetilde{\mathcal{U}}_{s}, \widetilde{G}_{s}, \widetilde{Z}, \widetilde{R}, \widetilde{\epsilon}_{s},  \widetilde{g}_{s}): \Omega \mapsto K([0,$ $ 1])  \times  \D(\X) \times \Z \times \{0,1\} \times \mathcal{E} \times l_{2}(\X, \Z_{1}, \mathcal{E}),  $
	where $ K([0,1]) $ is the collection of all non-empty closed subsets of the unit interval,  and $ \D(\X) $ denotes the space of distribution functions on $ \X. $  We say $ \Phi $	is the set of counterfactual experiments, if for all $ \phi \in \Phi $, we have  (i)  $ \widetilde{G}_{s}^{-1}(U_{s})  =  \widetilde{G}_{s}^{-1}(U'_{s}) $ almost surely for all $  U_{s}, U'_{s} \in \widetilde{\mathcal{U}}_{s} $; (ii) $ ( \widetilde{\epsilon}_{s} ,\widetilde{Z} ,\widetilde{R}) \stackrel{d}{=} (\epsilon_{s}, Z, R) $; (iii) $ \widetilde{g}_{s}  = g_{s} $, (iv) for all $  U_{s} \in \mathcal{U}_{s} $ and $  \widetilde{U}_{s} \in \widetilde{\mathcal{U}}_{s}  $,  there exists $  \widetilde{U}'_{s} \in \widetilde{\mathcal{U}}_{s}  $ and $  U'_{s} \in \mathcal{U}_{s} $, respectively, such that  $( \widetilde{U}'_{s}| \widetilde{Z}_{1}, \widetilde{R} = 1) \stackrel{d}{=}  ( U_{s}|  Z_{1}, R = 1) $ and $ ( \widetilde{U}_{s}| \widetilde{Z}_{1},\widetilde{R} = 1) \stackrel{d}{=}  ( U'_{s}| Z_{1}, R = 1)  $, where $  \mathcal{U}_{s}   =  \{  \breve{U} \in \mathcal{U}[0,1]  :   (F^{-1}_{X_{s}|R}(\breve{U}_{s}|1)| Z_{1}, R = 1) \stackrel{d}{=} ( X_{s}| Z_{1}, R =1 ) \} $. 
\end{definition}

The definition of counterfactual experiments does not  specify the counterfactural target covariate $ \widetilde{X}_{s} $ directly. It is implicitly defined through the first two elements of $ \phi $.  The first element, $ \widetilde{\mathcal{U}}_{s} $, is a set of rank variables associated with the counterfactual target covariate, $ \widetilde{X}_{s} $. When $  \widetilde{X}_{s}  $ is absolutely continuous, $  \widetilde{\mathcal{U}}_{s} $ becomes a singleton set, but the set is generally not degenerate when the distribution of $ \widetilde{X}_{s} $ contains a mass point.  The second component, $ \widetilde{G}_{s} $, is the counterfactual distribution of $ \widetilde{X}_{s}  $ conditional on the study population.  When the target covariate is continuously distributed,   $ \widetilde{G}_{s} $ is continuous and strictly increasing, and therefore, $ \widetilde{X}_{s} $ is uniquely determined by $ \widetilde{X}_{s}  = \widetilde{G}_{s}^{-1}(\widetilde{U}_{s}) $, where $ \widetilde{U}_{s} $ is the only element in $  \widetilde{\mathcal{U}}_{s}  $.   However, when the target covariate contains mass points, there is a set of counterfactual rank variables that correspond to the same target covariate in the study population. This equivalent class is defined by Condition (i).  

Following \citet{rothe2012partial}, we restrict our attention to counterfactual changes where only the marginal distribution of $ X_{s} $ is changed, while the marginal distribution of $ Z $ and the dependence structure between $ X_{s} $  and $ Z $ remain unaffected.   This notion of a \textit{ceteris paribus} change is formally characterized by Conditions (ii)--(iv).  Condition (ii) implies that the joint distribution of the observed variables $ (Z, R) $ and the latent variable $ \epsilon_{s} $ remain unchanged across counterfactual experiments. 
Under Condition (iii), the structural production function, $ g $ is also not affected by the counterfactual change.  Condition (iv) imposes a rank similarity condition. It says the conditional rank of the counterfactural target covariate follows the same distribution as the status quo.  Due to the possibility of multiplicity of rank variables, the condition is also framed in terms of a set equivalence condition.  When we restrict attention to absolutely continuous target covariates, both $ \mathcal{U}_{s} $ and $ \widetilde{\mathcal{U}}_{s} $ are singleton sets. Hence, this condition reduces to $ (\widetilde{X}_{s} | \widetilde{Z}_{1}, \widetilde{R} = 1)  \stackrel{d}{=}  (X_{s} | Z_{1}, R = 1) $.

Each counterfactual experiment   $ \phi $  represents a modification of the underlying economic system. It completely determines the counterfactual outcome in the study population.  Yet we remain largely agnostic as to the counterfactual change in the auxiliary population. The definition also leaves the mechanism causing the change in the marginal distribution of the target covariate unspecified.

\begin{remark}
	Our definition of counterfactual experiments relaxes the rank invariance conditions imposed by \citet{rothe2012partial}. Instead, counterfactual changes in our context only need to satisfy a rank similarity or copula invariance condition. 
\end{remark}



With the counterfactual experiments defined,  we now construct the counterfactual covariate vector by $ \widetilde{W}_{G} := ( \widetilde{G}_{s}^{-1}(\widetilde{U}_{s}), \widetilde{Z}_{1}')'. $ The counterfactual outcome of the study population is then defined as $ \widetilde{Y}_{s} = \widetilde{g}_{s}(\widetilde{W}_{G}, \widetilde{\epsilon}_{s}) $, which follows a marginal distribution, $ F_{\widetilde{Y}_{s}},$ and a conditional distribution restricted to the principal population, $  F_{\widetilde{Y}_{s}|R =1} $. Note that the unconditional distribution is not well-defined, due to the lack of information on counterfactual changes in the auxiliary population. Therefore, we focus exclusively on the counterfactual distribution conditional on the study population in what follows. When $ X $ is discrete, a single counterfactual experiment is mapped to a set of counterfactual outcomes, and we denote the corresponding set of counterfactual distributions by $ \F_{\widetilde{Y}_{s}} $.

In our context, the sequence of counterfactual distributions is defined in terms of the ``marginal'' distribution of the potential covariate $ X_{s} $ in the study population, rather than the true unconditional distribution of the observed $ X $. Although $ X_{s} $ is missing from the main dataset, and therefore, its marginal distribution cannot be directly identified from the study population, we show in Theorem \ref{thm.main} that it can be recovered from the auxiliary data under the rank similarity assumption we impose in Assumption \ref{assmp.data.struct}.

The policy parameter we seek to identify in this paper is the pathwise derivative of counterfactual distributional effect conditional on the study population. It is adapted from the definition of  the \textit{marginal partial distributional policy effect} (MPPE) by \citet{rothe2012partial}.

\begin{definition}[Marginal Partial Distributional Policy Effect]
	
	Let  $  \Phi^{\ast} := \{ \phi_{t} \}_{t\geq 0}  \subset \Phi $ denote a sequence of counterfactual experiments,  such that $\widetilde{G}_{s,t}  \to   F_{X_{s}|R =1} $, as $ t \downarrow 0 $.
	The MPPE for a given functional $ \nu:  \D(\Y)  \to \mathbb{R}$ and a sequence of $ \widetilde{F}_{s,t} \in \F_{\widetilde{Y}_{s,t}}  $  is defined by,
	\begin{equation*}
	MPPE(\nu, \{\widetilde{Y}_{s,t}\}_{t\geq 0} ) :=  \left. \dfrac{\partial \nu(F_{\widetilde{Y}_{s,t}|R=1})}{\partial t} \right\vert_{t= 0} = \lim_{t \downarrow 0} \dfrac{\nu(F_{\widetilde{Y}_{s,t}|R=1}) - \nu(F_{Y_{s}|R=1}) }{t}. 
	\end{equation*}
\end{definition}

We consider two specific types of counterfactual distributional changes: MDS and MQS.  The defintion of the former is due to \citet{firpo2009unconditional}. It denotes a small perturbation in the distribution of $ X_{s} $, in the direction of $ G $. MQS, on the other hand, considers a minuscule change in the quantiles of $ X_{s} $.  This type of policy change includes the MLS, $ G^{-1}_{t,ls}(u) :=  F_{X_{s}|R}^{-1}(u|1) + t $, as a special case.   

\begin{definition}[Counterfactual Policy Distributions] \ 
	\begin{itemize}
		\item Marginal Distributional Shift (MDS): $ G_{t,p}(x) := F_{X_{s}|R}(x|1) + t(G(x) - F_{X_{s}|R}(x|1)). $ 
		\item Marginal Quantile Shift (MQS): $ G^{-1}_{t,q}(u) :=  F_{X_{s}|R}^{-1}(u|1) + t(G^{-1}(u) - F_{X_{s}|R}^{-1}(u|1)).  $ 
	\end{itemize}
\end{definition}

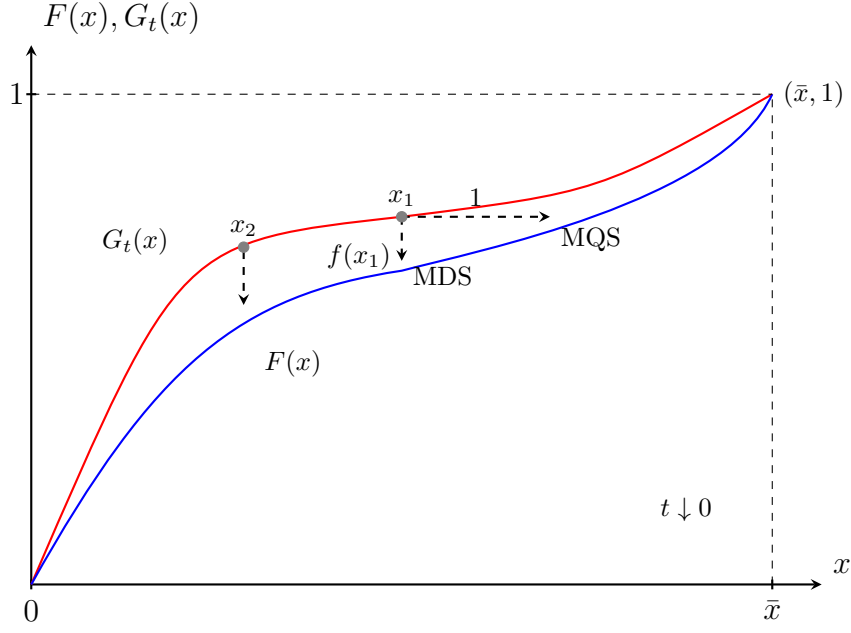
\begin{figure} 
	\centering 
	\begin{tikzpicture}[scale = 6.5] 
	
	\draw[thick, black, -stealth] (0, 0) -- (1.60, 0)
	node[anchor=south west] {$x$};
	\draw[thick, black, -stealth] (0, 0) -- (0, 1.1)
	node[anchor=south west] {$F(x), G_{t}(x)$};
	
	\draw[thick] (0, 0.01) -- (0, -0.01)
	node[anchor=north] {\(0\)};
	
	\draw[thick] (1.5, 0.01) -- (1.5, -0.01)
	node[anchor=north] {\(\bar{x}\)};
	
	\draw[thick] (-0.01, 1) -- (0.01, 1)
	node[anchor=east] {\(1\)};
	
	
	\draw[thick, red] (0, 0)
	.. controls (0.3, 0.7) and (0.3, 0.7) .. (0.75, 0.75)
	.. controls (1.15, 0.8) and (1.15, 0.8) .. (1.5, 1);

	\draw[thick,  blue] (0, 0)
	.. controls (0.25, 0.45) and (0.42, 0.59) .. (0.75, 0.64)
	.. controls (0.9, 0.68) and (1.4, 0.78) .. (1.5, 1);

	\draw[thick, dashed, -stealth] (0.75, 0.75) -- (0.75, 0.66);
	\node[below right] at (0.75, 0.67) {\footnotesize MDS};
	\node[ left] at (0.75, 0.67) {\footnotesize $f(x_{1})$};
	\draw[thick, dashed, -stealth] (0.75, 0.75) -- (1.05, 0.75);
	\node[below right] at (1.05, 0.75)  {\footnotesize MQS};
	\node[above] at (0.9, 0.75)  {\footnotesize 1};
	
	\draw[thick, dashed, -stealth] (0.43, 0.68) --(0.43, 0.57);

	\node[above left] at (0.3, 0.65) {\footnotesize $G_{t}(x)$};
	\node[below right] at (0.45, 0.5) {\footnotesize $F(x)$};
	\node[below left] at (1.4, 0.2) {\footnotesize $ t \downarrow 0 $};
	
	\filldraw [gray] (0.75, 0.75) circle (0.3pt);
	\node[above] at (0.75, 0.75) {\footnotesize $ x_{1} $};
	
	\filldraw [gray] (0.43, 0.688) circle (0.3pt);
	\node[above] at (0.43, 0.688) {\footnotesize $ x_{2} $};
	
	\draw [dashed] (0, 1) -- (1.5,1)
	node[anchor= west] {\footnotesize $(\bar{x}, 1)$};
	\draw [dashed] (1.5, 0) -- (1.5, 1);
	\end{tikzpicture}
	\caption{ {Marginal Distributional Shift and Marginal Quantile Shift}}
	\label{fig: mds.mqs}
\end{figure}

\begin{remark}
	Figure \ref{fig: mds.mqs} illustrates how the rates of change between the two types of counterfactuals are related. Under the condition that $F_{X_{s}|R = 1} $ is compactly supported with strictly positive density on $ \X $,  MQS in the direction of $ q(x) $ can be approximated in the limit by MDS with $ G(x) = F_{X_{s}|R}(x|1) - f_{X_{s}|R}(x|1) q(x)$.  
\end{remark}
Turning to the case of quantiles,   the quantile operator for a particular $ \tau $ is defined by,  $ \nu_{\tau}(F_{Y_{s}|R =1}) := F_{Y_{s}|R= 1}^{-1}(\tau) $.  With the understanding that MPPE associated with a counterfactual experiment is generally a set when the X is discretely valued, we suppress the index with respect to $ \{\widetilde{Y}_{s,t}\}_{t\geq 0} $ for notational convenience, and denote the MPPE with MDS, $ MPPE(\nu_{\tau}, \{\widetilde{Y}_{s,t}\}_{t\geq 0}   )   $, and MPPE with MQS, $ MPPE(\nu_{\tau}, \{\widetilde{Y}_{s,t}\}_{t\geq 0} )  $, by $ UQE_{p}(\tau, G) $ and $ UQE_{q}(\tau, G)$, respectively. 
Here, and in what follows, the qualifier ``unconditional'' in UQE should be understood as conditional on (or relative to) the study population.

\subsection{Identification of $ F_{Y_{s}|X_{s}Z_{1}R = 1} $} \label{sec: Fyw}
If there is no missing variable, the joint distribution of $ (Y, X, Z) $ is directly identifiable from a random sample. Under data combination, however, only the  ``marginal'' conditional distributions: $ F_{Y_{s}|Z R = 1} $ and $ F_{X_{a}|Z R = 0}  $, can still be separately identified from the two samples, respectively. The conditional distribution, $ F_{Y_{s}|X_{s}Z_{1}R = 1} $, is generally not identifiable without further cross-population assumptions. 

Instead of seeking identification of the entire conditional distribution,\footnote{Information of the entire distribution is crucial for the conditional quantile regression approach, even if the researcher may be interested in a single fixed quantile.} $ F_{Y_{s}|X_{s}Z_{1}R = 1}(\cdot|\cdot, $ $\cdot , 1) $, we demonstrate in Section \ref{sec: cont.ex} and \ref{sec: bin.ex} that UQE, and MPPE in general, can be identified using information at a single point of the distribution.  This allows us to obtain identification under much milder restrictions on the pseudo-merged population.  Identification is achieved through the excluded instrument variables, $ Z_{2} $.

To ease notational burden, define $   \Lambda(x, z_{1})  := F_{Y_{s}|X_{s} Z_{1} R }( q_{\tau}| x, z_{1}, 1) $ for all  $ x\in \X$ and $z_{1} \in \Z_{1} $. 

\begin{assumption} \label{assmp.ident.fyw.1}
	$ \epsilon_{s} \independent Z_{2} | X_{s}, Z_{1}, R =1. $
\end{assumption}
Assumption \ref{assmp.ident.fyw.1}  implies that $ Z_{2} $ can be excluded from the outcome equation, and therefore, can be used as a source of exogenous variation to proxy for the missing covariate in the study population.  Note that $ \Lambda $ is generally not identified without an exogenous instrument $ Z_{2}$. We illustrate this point with the example below.

Under Assumptions \ref{assmp.data.struct} and \ref{assmp.ident.fyw.1},  the following moment-matching equation holds, for all $ z \in \mathcal{Z} $,
\begin{equation}\label{eq.ident.fyw}
\mathbb{E}\left[1(Y \leq q_{\tau})| Z, R = 1\right] -   \mathbb{E}\left[\Lambda(W)  | Z,R =0 \right] = 0.
\end{equation}

Equivalently, $ \Lambda $ can be identified based on a likelihood-ratio-weighting equation,
\begin{equation}\label{eq.ident.fyw.1}
\mathbb{E}\left[\left. R1(Y \leq q_{\tau}) -   (1-R)\Lambda(W) \dfrac{r(Z)}{1-r(Z)}  \right\vert Z \right] = 0.
\end{equation}

The next assumption is about the global identification of $ \Lambda $. 
\begin{assumption}\label{assmp.ident.fyw.2} 
	$ \Lambda $ is the unique solution to \eqref{eq.ident.fyw} or \eqref{eq.ident.fyw.1} almost surely. 
\end{assumption}

Assumption \ref{assmp.ident.fyw.2} is a high level condition. It is implied by a bounded completeness condition.\footnote{For two random element $ U $ and $ V $, we say $ U $ is bounded complete for $ V $, relative to  a subpopulation $ S = s $, if  for all bounded measurable functions $ \delta(\cdot)$, $ \E[\delta( U) |V, S = s ] = 0 $ implies $ \delta(U) \equiv 0 $ almost surely.} Note that $ \Lambda $ is globally identified as long as $ \E[ \Lambda(W) - \widetilde{\Lambda}(W) |Z, R = 0]  = 0$ implies $ \Lambda  = \widetilde{\Lambda} $, which follows immediately if $ \Lambda $ is  measurable with respect to $ W $, and  $ W $ is bounded complete for $ Z $, relative to the auxiliary population.  Bounded completeness is weaker than the commonly adopted completeness condition appearing in \citet{newey2003instrumental} and \citet{fan2014identifying}.  We refer readers to \citet{hoeffding1977some}, \citet{blundell2007semi}, and \citet{lehmann2012statistical} for detailed discussions. 

In Section \ref{estimation}, we base our estimation and inference on parametric identification of $ \Lambda $. In this case, we assume that $ F_{Y_{s}|X_{s} Z_{1} R }( q_{\tau}| x, z_{1}, 1)  = \Lambda(x, z_{1}; \beta_{0}) $, for $ \beta_{0} \in \theta_{\beta} \subset \mathbb{R}^{d_{\beta}} $. In Lemma \ref{lemma.bdd.comp},  we provide a set of sufficient conditions which allow us to establish a global parametric identification condition analogous to Assumption \ref{assmp.ident.fyw.2}. 

\begin{lemma}\label{lemma.ident.fyw}
	$ F_{Y_{s}|X_{s}Z_{1}R}(q_{\tau}|\cdot, \cdot, 1) $ is point identified under Assumptions \ref{assmp.data.struct}--\ref{assmp.ident.fyw.2}. 
\end{lemma}

Lemma \ref{lemma.ident.fyw} establishes the nonparametric identification of $ \Lambda $.  The proof for the parametric case follows along exactly the same line so we omit it here.  In the next example, we verify the identification assumptions in a conditional normal model.  

\begin{example}[Conditional Normal Model]\label{eg.1}
	Let the structural equations of the study population be given by 
	\begin{align*}
	Y_{s}  & = g_{s}( X_{s}, Z_{1}) + \epsilon_{s}, \\
	X_{s}  & = h_{s}(Z)  + \eta_{s},
	\end{align*}
	where $ \epsilon_{s} $ and $ \eta_{s} $ are jointly normally distributed. Specifically,  for positive-valued functions $\psi_{y}(\cdot) $ and $ \psi_{x}(\cdot) $, we have
	\begin{equation*}
	(\epsilon_{s}, \eta_{s}) | Z,  R = 1 \stackrel{p}{\to}  N\left(0,   \begin{pmatrix}
	\psi_{y}(Z_{1})	& 0 \\ 0 & \psi_{x}(Z_{1})
	\end{pmatrix}  \right).
	\end{equation*}
Then,  $ \Lambda(w)  = \Phi(\psi_{y}(z_{1})^{-1/2} (q_{\tau} - g_{s}(w )) ) $, where $ \Phi(\cdot ) $ denotes the CDF of standard normal distribution.
Suppose the reduced-form of $ X $ given $ Z $ in the auxiliary population is $ 	X_{a}  = h_{a}(Z) + \eta_{a}, $ Assumption \ref{assmp.data.struct}(e) is satisfied if $ h_{s} =  h_{a}  = h $, and $( \eta_{s} | Z, R = 1) \stackrel{d}{=} (\eta_{a} | Z, R = 0 )$.  Assumption \ref{assmp.ident.fyw.1} holds if $ Z_{2} \independent \epsilon_{s} | X_{s}, Z_{1}, R = 1 $.  Assume, in addition that, conditional on $ z_{1} $,  $ \supp(Z_{2}) $  contains an open set and that $ h(z_{1}, \cdot) $ maps open sets of $ z_{2} $ into open sets. Assumption \ref{assmp.ident.fyw.2} then follows by Theorem 2.2 in \citet{newey2003instrumental}.

Turning to the linear case,  let  $ g_{s}(w) = \gamma_{s_{1}} x + \gamma_{s_{2}}'z_{1}  $,  $  h(z)= \delta_{1}' z_{1} + \delta_{2}' z_{2}   $, $ \psi_{y} =  \psi_{x}  = 1$,  $ \eta_{s} \independent (\epsilon_{s},  Z_{2}) $, and therefore, $ \mathbb{E}\left[1(Y \leq q_{\tau})| Z, R = 1\right] =  \Phi((q_{\tau}-  (\gamma_{s_{1}} \delta_{1}' + \gamma_{s_{2}})'Z_{1} - \gamma_{s_{1}} \delta_{2}'Z_{2} )/(1+\gamma_{s_{1}}^{2})^{1/2}). $ As a consequence, $ (\gamma_{s_{1}}, \gamma_{s_{2}}')' $ are uniquely determined by \eqref{eq.ident.fyw} or \eqref{eq.ident.fyw.1}, if and only if $ \delta_{2} \neq 0$. 

\end{example}

\subsection{Identification with Continuously Distributed $ X $} \label{sec: cont.ex}

In this section, we  establish the identification  of $ UQE_{q} $ and $ UQE_{p} $ when the distribution of $ X $ is absolutely continuous. Before stating the main result, we need some additional identifying  assumptions. 

\begin{assumption}\label{assmp.ident.ex.cont}\
	\begin{enumerate}[label=(\alph*)]
		\item  (i) $ \epsilon_{s} \independent U_{s} | Z_{1}, R = 1$; (ii) there exists a $ t_{0} $ sufficiently close to 0, such that for all $ t \leq t_{0} $ and  $  \phi_{t} \in \Phi^{\ast} $,  $ \widetilde{\epsilon}_{s,t} \independent \widetilde{U}_{s,t}| \widetilde{Z}_{1,t}, \widetilde{R}_{t} = 1 $.  \label{ident.ex.cont.indep}
		\item $ \supp(G) \subset \supp(F_{X|Z R}(\cdot |Z, 1)) $ almost surely. \label{ident.ex.cont.supp}
	\end{enumerate}
\end{assumption}

\begin{assumption} \label{assmp.ident.hd}
	$ F_{Y|R = 1}  $ is continuously differentiable in an open neighborhood of $ q_{\tau}$ with strictly positive density function $ f_{Y|R = 1} $.
\end{assumption}

Assumption \ref{assmp.ident.ex.cont}(a)(i) says that conditional on $ Z_{1} $, structural error $ \epsilon $ is independent of the rank variable $ U_{s} $ in the study population. This is  much weaker than the commonly assumed strict independence condition that $ X_{s} $ is independent of $ \epsilon_{s} $ unconditionally.  Conditional exogeneity has also been imposed by \citet{firpo2009unconditional},  \citet{rothe2012partial}, and \citet{chernozhukov2013inference}, among others. 
Assumption \ref{assmp.ident.ex.cont}(a)(ii) requires the conditional independence condition of  part (a) to hold when counterfactual experiments get sufficiently close to the status quo. Under the rank invariance condition imposed by \citet{rothe2012partial}, it is automatically implied by Assumption \ref{assmp.ident.ex.cont}(a)(i). 
Assumption \ref{assmp.ident.ex.cont}(b) ensures that the conditional distribution of $ Y_{s} $ given $ W $  is identified over the support of $ W $. 
Assumption \ref{assmp.ident.hd} imposes a smoothness condition on the distribution of target outcome, which implies that $ F_{Y|R = 1}^{-1} $ is Hadamard differentiable at $ F_{Y|R = 1}  $, tangentially to the set of functions that are continuous at  $ q_{\tau} $. 

The main theoretical result of this section is given as follows. 

\begin{theorem} \label{thm.main}
	Suppose that Assumptions \ref{assmp.data.struct}--\ref{assmp.ident.hd} hold,  and that the distribution of $ X $ is absolutely continuous with respect to the Lebesgue measure,  both $UQE_{p}(\tau, G) $ and $UQE_{q}(\tau, G) $ are identified. 
	
	(a)  For $ UQE_{q} $, we have
	\begin{equation*}
	UQE_{q}(\tau, G) =  - \dfrac{1}{f_{Y_{s}|R}(q_{\tau}|1)(1-Q_{0})} \mathbb{E}\left[ (1-R)\ell(Z) \Lambda_{x}(X, Z_{1}) g_{q}(X)  \right],
	\end{equation*}
	where $ g_{q}(x) := G^{-1}(F_{X_{s}|R}(x|1)) - x  $, and $ F_{X_{s}|R}(x |1) = \frac{1}{1-Q_{0}}\mathbb{E}\left[  (1-R )\ell(Z) 1(X\leq x) \right]$.
	
	(b) Suppose in addition that $ \X $ is compact, and  $ F_{X_{s}|R = 1}  $ is continuously differentiable on $ \X $ with strictly positive density function $ f_{X_{s}|R = 1} $. Then we have,
	\begin{equation*}
	UQE_{p}(\tau, G) = - \dfrac{1}{f_{Y_{s}|R}(q_{\tau}|1)(1-Q_{0})}\mathbb{E}\left[ (1-R)\ell(Z)  \Lambda_{x}(X, Z_{1})  g_{p}(X)   \right],
	\end{equation*}
	where $ g_{p}(x) := -  \frac{ G( x ) - F_{X_{s}|R}(x|1) }{f_{X_{s}|R}( x |1 )}$, and $ f_{X_{s}|R}( x |1 ) =  \frac{1}{1-Q_{0}}\partial \mathbb{E}\left[  (1-R )\ell(Z) 1(X\leq x) \right] /\partial x.$
\end{theorem}

\begin{remark}
	The compactness condition on $ \X $  is assumed to ensure the existence of pathwise derivative of the inverse map. It can be relaxed by imposing a boundary condition on $  \Lambda_{x}  $. Specifically, we may assume that $  \Lambda_{x} $ vanishes when $ x \not\in [F_{X_{s}|R=1}(q_{1}) + \epsilon,  F_{X_{s}|R=1}(q_{2}) - \epsilon  ]  $,  for $ 0<q_{1} < q_{2} < 1$ and some $  \epsilon >0 $.
\end{remark}

\subsection{Identification with Discrete Covariate} \label{sec: bin.ex} 

Let the support of $ X $ be $ \{x^{1},\dots, x^{l} \} $. When $ X $ is discrete,  MQS is not well-defined and we consider  MDS only, with counterfactual experiments defined through a fixed discrete distribution, $ G $.  

\begin{assumption}\label{assmp.bin} \ 
	\begin{enumerate}[label=(\alph*)]
		\item  (i) $ \epsilon_{s} \independent U_{s} | Z_{1}, R = 1$, for all $ U_{s} \in \mathcal{U}_{s}  $; (ii)	there exists a $ t_{0} $ sufficiently close to 0, such that for all $ t \leq t_{0} $ and  $  \phi_{t} \in \Phi^{\ast} $,   $ \epsilon_{s,t} \independent \widetilde{U}_{s,t}| \widetilde{Z}_{1,t}, \widetilde{R}_{t} = 1$,  for all $   \widetilde{U}_{s,t} \in \mathcal{\widetilde{U}}_{s,t}$.
		\item  $ \supp(G) \subset \supp(F_{X_{s}|R = 1}). $
		\item  For all $ U_{s} \in \mathcal{U}_{s} $, $ F_{U_{s}|Z_{1}R}(u_{s}|z_{1}, 1) $ is continuously differentiable in $ u_{s} $, for all $ z_{1} \in \Z_{1} $. 
	\end{enumerate}
\end{assumption}

Assumption \ref{assmp.bin}(a) is the counterpart of Assumption \ref{assmp.ident.ex.cont}(a) for discrete covariates.  Since the rank variables are no longer uniquely pinned down by strictly increasing quantile functions, we  strengthen Assumption \ref{assmp.ident.ex.cont}(a) so that conditional independence holds for all the rank variables in the equivalent class.  With this identifying assumption in hand, we are ready to present the following identification result.  For $ j = 1,\dots,l $, let the period bound generating function be defined by
\[h_{q_{\tau}}(x^{j}, x^{j-1}, z_{1}) := - \dfrac{ (\Lambda(x^{j-1}, z_{1})  - \Lambda(x^{j}, z_{1}) ) \cdot    ( G(x^{j-1}) -   F_{X_{s}|R}(x^{j-1}|1) ) }{f_{Y_{s}|R}(q_{\tau}|1)} . \]

\begin{theorem} \label{thm.main.discrete}
	Suppose that Assumptions \ref{assmp.data.struct}--\ref{assmp.ident.fyw.2}, \ref{assmp.ident.hd}, and \ref{assmp.bin} hold, $ UQE_{p}(\tau, G)  $ is partially identified, with
	\begin{multline*}
	UQE_{p}(\tau, G)   \in \left[   \sum_{j \in \J_{+}}  h_{q_{\tau}}(x^{j}, x^{j-1}, z^{\dagger}_{1,j}) +  \sum_{j \in \J_{-}}  h_{q_{\tau}}(x^{j}, x^{j-1}, z^{\ast}_{1,j}), \right. \\
	\left.  \sum_{j \in \J_{+}}  h_{q_{\tau}}(x^{j}, x^{j-1}, z^{\ast}_{1,j}) +  \sum_{j \in \J_{-}}  h_{q_{\tau}}(x^{j}, x^{j-1}, z^{\dagger}_{1,j}) \right], 
	\end{multline*}
	where  $ \J_{+}  := \{ j \in \{1,\dots,l\}:    G(x^{j-1}) \leq   F_{X_{s}|R}(x^{j-1}|1)   \}$ ($ \J_{-}  $ is analogously defined), 
	$  z_{1,j}^{\ast} := \arg\sup_{z_{1} \in \Z_{1}} (\Lambda(x^{j-1}, z_{1})  - \Lambda(x^{j}, z_{1}))$,  $  z_{1,j}^{\dagger} := \arg\inf_{z_{1} \in \Z_{1}} (\Lambda(x^{j-1}, z_{1})  - \Lambda(x^{j}, z_{1}))$, and $ F_{X_{s}|R }(x^{j}|1) = \E[ \frac{1-R}{1-Q_{0}} \ell(Z) 1(X \leq x^{j})  ] $, for  $ j \in \{1,\dots, l \} $. 
	
\end{theorem}

Theorem \ref{thm.main.discrete} indicates that $ UQE_{p} $ is generally partially identified with bounds generated by $ h_{q_{\tau}}. $  In the special case when $   \Lambda(x, z_{1})  $ is constant in $ z_{1} $, the identified set of $  UQE_{p}  $ reduces to a singleton.    

If $ X $ is binary and $ G_{t, p}(x) = 1\{ 0\leq x < 1 \} (F_{X_{s}|R}(0|1) - t) + 1\{x \geq 1\} $,  $ h_{q_{\tau}} $ reduces to $  -  \left( \Lambda(1, z_{1}) - \Lambda(0, z_{1}) \right) /f_{Y|R}(q_{\tau}|1)   $. In such circumstance, Theorem \ref{thm.main.discrete} corresponds to the two-sample generalization of Theorem 5 in \citet{rothe2012partial}, when $ \nu $ in that paper takes on the quantile functional.

\section{Estimation and Inference} \label{estimation}

In this section, we  discuss estimation and inference for our two-sample UQE.  First, we describe an estimation procedure for $ UQE_{q} $ and $ UQE_{p} $ as identified in Theorem \ref{thm.main}.  We then show that our estimator is consistent and asymptotically normal in Theorem \ref{thm.asy.nor}.\footnote{Here we focus on the scenario where the distribution of $ X $ is absolutely continuous.   When $ X $ is discrete, the problem features partially identified parameters defined by the intersection bounds.  \citet{chernozhukov2013intersection} provide an extensive treatment of this topic. We omit discussion here and refer readers to Appendix D in \citet{rothe2012partial}  for a detailed discussion on how to apply their method. }

\subsection{Estimation Procedure}
Following the discussion in Section \ref{identification},  we first propose an estimator of the conditional probability, $ \Lambda $. Here, we restrict our attention to the parametric setting where $ \Lambda $ is indexed by a vector of parameter, $ \beta $.   We use a moment-matching method based on \eqref{eq.ident.fyw.1} to estimate $ \beta $.  The estimation of $ \beta $ consists of four-steps. In the first step, we estimate $ q_{\tau} $ by solving
\begin{equation}\label{est.step1}
\widehat{q}_{\tau} := \arg\min_{q \in \Y}\mathbb{E}_{n}[R(\tau - 1(Y \leq q))\cdot (Y - q)]  .
\end{equation} 
The next three steps follow closely the Auxilliary-to-Study Tilting (AST) method proposed by \citet{graham2016efficient}. Using the AST estimator, $ \beta $ and the propensity score can be jointly estimated from moment restrictions in  \eqref{eq.ident.fyw.1}.  To implement the estimator, we first estimate the propensity score, $ r(z) $.  Towards this end, we assume that the propensity score takes a parametric form, ie. $  r(z) = L( k(z)'\gamma) $, where $ L(\cdot) $ is any link function that satisfies Assumption \ref{assmp.cr.beta}(e).  Using $  L(\cdot) $,   $ \widehat{\gamma} $  can be obtained by solving  the following problem,
\begin{equation}
\widehat{\gamma} := \arg\max_{\gamma \in \Theta_{\gamma} } \E_{n} \left [R \log( L( k(Z)'\gamma) ) + (1- R) \log( 1- L( k(Z)'\gamma) ) \right]. \label{est.step2}
\end{equation}
 The AST estimator augments the conditional maximum likelihood estimator $ \widehat{\gamma} $ with tilting parameters. The resulting estimator of $ \beta $ is more efficient than the one based on $ \widehat{ \gamma } $ alone.
Let $ t(z) $ be a vector of known functions of $ z $ with a constant term as the first element.  Denote the tilting parameters associated with the auxiliary data and the study sample, by $ \lambda_{a} $ and $ \lambda_{s} $, respectively. They are estimated by solving, 
\begin{align}
\E_{n} & \left[ \left(  \dfrac{1-R}{  1 - L( k(Z)'\widehat{\gamma}  + t(Z)' \widehat{\lambda}_{a}  )  } - 1  \right)  L(k(Z)' \widehat{ \gamma }) t(Z) \right] = 0, \label{est.step3.1} \\
\E_{n} & \left[ \left(  \dfrac{ R}{  L( k(Z)'\widehat{\gamma}  + t(Z)' \widehat{\lambda }_{s} )  } - 1  \right)  L(k(Z)' \widehat{ \gamma }) t(Z) \right]= 0. \label{est.step3.2}
\end{align}
Using $ \widehat{ \lambda}_{s} $ and $ \widehat{ \lambda}_{a}  $,  we compute study and auxiliary sample tilts, which are defined as follows
\begin{align}\label{est.step3.3}
\widehat{\pi}_{i}^{s}  &:=  \dfrac{   L( k(Z_{i})'\widehat{\gamma} )  }{   L( k(Z_{i})'\widehat{\gamma}  + t(Z_{i})' \widehat{\lambda}_{s})  }, \qquad \qquad \widehat{\pi}_{i}^{a}  :=  \dfrac{   L( k(Z_{i})'\widehat{\gamma} )  }{  1 - L( k(Z_{i})'\widehat{\gamma}  + t(Z_{i})' \widehat{\lambda}_{a} )  }.
\end{align}

Also let $ e(z) $ be a $ d_{\beta} $-dimensional vector of known functions of $ z $, and $ g( a; \widehat{q}_{\tau}, \widehat{\gamma}, \widehat{\lambda}_{s}, \widehat{\lambda}_{s}, \beta )$ $ :=   ( \widehat{\pi}^{s} r 1(y\leq \widehat{q}_{\tau})  - \widehat{\pi}^{a} (1-r)\Lambda(w;\beta)) e(z) $.    Now, in the last step,  $ \beta $ can be estimated by
\begin{equation} \label{est.step4}
\widehat{\beta} := \arg\inf_{\beta \in \Theta_{\beta} } \widehat{\LL}_{n}(\beta), 
\end{equation}
where $\widehat{\LL}_{n}(\beta)   :=  \norm{\E_{n}[g(A; \widehat{q}_{\tau}, \widehat{\gamma}, \widehat{\lambda}_{s}, \widehat{\lambda}_{a}, \beta ) ]}^{2}_{\Omega_{n}} $ and $ \norm{x}^{2}_{\Omega_{n}} := x'\Omega_{n}x   $, for a sequence of positive definite weighting matrices $ \Omega_{n} $.

Using these quantities, we can obtain  $ \Lambda_{x}(W;\widehat{\beta})  := \partial  \Lambda(W;\widehat{\beta})/\partial x $,  and $ \widehat{\ell }(z) := \frac{n_{a}}{n_{s}} \cdot \frac{L(k(z)'\widehat{ \gamma } + t(z)' \widehat{\lambda}_{s} )}{ 1- L(k(z)'\widehat{ \gamma } + t(z)' \widehat{\lambda}_{a} )}. $
Throughout this section, we assume that the counterfactual distribution $ G $ is known.  In practice, if $ G $ is not known, it may be estimated from an independent sample; see e.g. \citet{rothe2010nonparametric}.   Using the above estimates and $ \widehat{F}_{X_{s}|R = 1}(\cdot) := \E_{n_{a}}[\widehat{\ell }(z) 1(X \leq \cdot)] $, where $ \E_{n_{a}}[X]  $ denotes $   n_{a}^{-1} \sum_{i = n_{s}+1}^{n} X_{i}  $, $ \widehat{ g}_{q}  $ can be obtained as the plug-in estimator.   
For  $ g_{p} $, we need an estimator for $ f_{X|R=1}(\cdot) $. Our identification relies on a compact support condition, and it is well known that the Prazen-Rosenblatt density estimator is not valid near the boundary of support.  To overcome this challenge, we introduce trimming.\footnote{ Trimming is widely adopted  in the literature; see e.g. \citet{hardle1989investigating}, \citet{powell1989semiparametric} among others.   This specific trimming function is inspired by \citet{guerre2000optimal} and \citet{li2002structural}.  As an alternative, we can use a local polynomial density estimator that adjusts for the boundary bias adaptively; see \citet{cattaneo2020simple} for details.} For a kernel $ K_{x} $ with compact support, and some bandwidth $ b_{x} $, we let 
\begin{equation*}
\widehat{f}_{X|R }(x|1)  := \E_{n_{a}}\left[ \widehat{\ell}(Z) I_{b_{x}} K_{b_{x}}\left( X - x \right)   \right] ,
\end{equation*}
where  $  K_{b_{x}}(\cdot )  := b_{x}^{-1} K_{x}(\cdot/b_{x} )$.  $ I_{b_{x}} $ is a trimming indicator, which equals one for $x \in  \{  [ \ubar{x} + \rho_{x} b_{x}/2,   \bar{x} - \rho_{x} b_{x}/2 ] \}$, where $ \ubar{x} $, $ \bar{x} $, and $  \rho_{x} $ are the lower and upper bound, of $ \X $, and the diameter of $ \supp(K_{x}) $, respectively. The density, $ f_{Y|R } $, can also be estimated using kernel density estimator. Specifically,  for any kernel function $ K_{y}(\cdot) $ that satisfies Assumption \ref{assmp.uqe}(b),  let $ \widehat{f}_{Y|R}( y |1)  := \E_{n_{s}}[ K_{b_{y}}\left( Y_{i}- y\right)],  $ where $ \E_{n_{s}}[X] :=  n_{s}^{-1} \sum_{i = 1}^{n_{s}} X_{i}   $ and  $ K_{b_{y}}(y)  := b_{y}^{-1} K_{y}(y/b_{y})  $.  

Now, plugging in the estimators of nuisance quantities,  $ UQE_{j}(\tau, G)  $ can thus be estimated by,
\begin{equation} \label{est.step5}
\widehat{UQE}_{j}(\tau, G) := - \dfrac{1}{\widehat{f}_{YR}(\widehat{q}_{\tau}|1)} \mathbb{E}_{n_{a}} [\widehat{\ell}(Z)\Lambda_{x}(W;\widehat{\beta}) \widehat{g}_{j}(X)], \ j = p, q.
\end{equation}
We summarize the estimation procedure in the following algorithm.
\begin{algorithm}[Plug-in Estimator for $ \widehat{UQE} $] \ 
	\begin{enumerate}
		\item  Compute  the empirical quantile estimator  $ \widehat{q}_{\tau} $ by solving \eqref{est.step1}.
		\item  Compute  the conditional maximum likelihood estimator $ \widehat{ \gamma } $ by solving \eqref{est.step2}.
		\item  Solve \eqref{est.step3.1} and \eqref{est.step3.2} to get $ \widehat{ \lambda}_{j} $, and use them to compute $ \widehat{\pi}_{j} $, for $ j = s,a$, following \eqref{est.step3.3}.
		\item  Use the above quantities to compute $ \widehat{\beta} $, by solving \eqref{est.step4}.
		\item Compute $ \Lambda_{x}(\cdot; \widehat{\beta}), \widehat{\ell}(\cdot), \widehat{F}_{X_{s}|R = 1}(\cdot)$, $\widehat{f}_{X_{s}|R = 1}(\cdot)$. Using these quantities to compute  $ \widehat{ g}_{j}$, for $ j = p,q $. 
		\item For $ j  = p,q,$ compute the plug-in estimator $ \widehat{UQE}_{j} $,  following \eqref{est.step5}.
	\end{enumerate}
\end{algorithm}

\subsection{ Large Sample Results}

In this section, we present inference results for the estimators introduced in the previous section.  We first establish large sample properties of $ \widehat{\beta} $, for which purpose,  some additional regularity conditions are in order.

\begin{assumption}  \label{assmp.cr.beta} \ 
	\begin{enumerate}[label=(\alph*)]
		\item  (i)$ \{ (R_{i}, R_{i}Y_{i}, (1-R_{i})X_{i}, Z_{i}) \}_{i = 1}^{n} $ are i.i.d.; (ii)  let $ \theta := (\gamma, \lambda_{s}, \lambda_{a}, \beta) \in \Theta :=  \Theta_{\beta}  \times \Theta_{\lambda}^{2} \times \Theta_{\beta} $, then $ \Theta $ is compact, and $ \theta_{0} $ lies in the interior of $ \Theta $.
		\item  $ F_{Y|Z R = 1}(y| z) $ is absolutely continuous and differentiable in $ y \in \Y_{0} $ for all $ z \in \Z $, where $ \Y_{0} $ is a compact subset of $ \Y $, and 
		\[ \sup_{(y,z) \in \Y_{0} \Z } | f_{Y|ZR}(y|z, 1)| \leq c_{1} < \infty.  \]    
		\item  (i) $ \Lambda(w;\beta) $ is twice continuously differentiable in $ \beta $ with uniformly bounded derivatives, for all $ w \in \W $;  
		(ii) $0 \leq \inf_{w,\beta}\Lambda(w;\beta),   \sup_{w,\beta}\Lambda(w;\beta) \leq 1$;  
		(iii) $ \Lambda_{x}(\cdot;\beta) $ is continuously differentiable in $ \beta $, and $ \sup_{w, \beta } |\Lambda_{x}(w;\beta) | \leq c_{2} <\infty $.
		\item  There exists a symmetric, non-random matrix $ \Omega $, such that  $  || \Omega_{n} - \Omega ||  = O_{p}(\delta_{\omega, n})$, where $ \delta_{\omega, n} = o(1) $, and that  $ c_{3}^{-1} \leq \lambda_{min}(\Omega)  \leq \lambda_{max}(\Omega) \leq c_{3} $.
		\item There is a unique $ \gamma_{0}  \in \Theta_{\gamma} $, and known function $ L(\cdot) $ such that (i) 
		\[ \ell(z) = \dfrac{1-Q_{0}}{Q_{0}} \cdot \dfrac{L(k(z)'\gamma_{0})}{1-L(k(z)'\gamma_{0})}. \]
		(ii) $ L(\cdot) $ is strictly increasing, twice continuously differentiable, with  bounded first and second order derivatives; (ii) $ \lim_{x \to -\infty} L(x) = 0 $ and $ \lim_{x \to \infty} L(x) = 1 $;  (iii)  $ 0 <c_{4} <L(k(z)' \gamma + t(z)'\lambda_{j} ) \leq c_{5} < 1 $ for all $ (\gamma, \lambda_{j}) \in \Theta_{\gamma} \times \Theta_{\lambda} $, $ j = s,a, $  and $ z \in \mathcal{Z} $. 
		\item  $  \E[|| j(Z)||^{4} ]  < \infty $, where $ j = k, t, e  $.
	\end{enumerate}
\end{assumption}

Assumption \ref{assmp.cr.beta}(a) is standard in the microeconometric literature.  Assumption \ref{assmp.cr.beta}(b)  requires the conditional density $ f_{Y|ZR}(\cdot | \cdot, 1) $ be bounded uniformly for all $ (y,z) \in \Y_{0} \Z $. 
Assumption \ref{assmp.cr.beta}(c) imposes mild smoothness conditions on the parametric function $ \Lambda(\cdot, \cdot  ;\cdot)  $, requiring it to be bounded between the unit interval, thus behaving like a distribution function. Assumption \ref{assmp.cr.beta}(d) states that $  \Omega_{n} $ is consistent for $ \Omega $, which is positive definite. Assumption \ref{assmp.cr.beta}(e) implies that the true ``propensity score'' is known up to finite dimensional $ \gamma_{0} $. It also specifies smoothness and boundedness conditions on the parametric propensity score. 
Finally, due to the estimation of $ q_{\tau} $,  we impose a finite fourth moment condition in Assumption \ref{assmp.cr.beta}(f), which is stronger than the usual square-integrability condition.

\begin{lemma} \label{lemma.beta}
	Suppose that Assumptions \ref{assmp.data.struct}--\ref{assmp.ident.hd}, and Assumption \ref{assmp.cr.beta} hold, then (i) $ \widehat{\beta}  \stackrel{p}{\rightarrow} \beta_{0}  $;  furthermore, (ii) suppose that the Jacobian, $ M_{\Omega} $, as defined in the Supplementary Appendix A, is invertible, then
	\begin{equation*}
	\sqrt{n} (\widehat{\beta} - \beta_{0})  = \dfrac{1}{\sqrt{n}} \sum_{i = 1}^{n} \psi_{\beta}(A_{i}; \theta_{0}, q_{\tau}) + o_{p}(1),
	\end{equation*}
	where $ \psi_{\beta}(A; \theta_{0}, q_{\tau}) $ is given in the Supplementary Appendix A, and  (iii)
	\begin{equation*}
	\sqrt{n} (\widehat{\beta} - \beta_{0})   \stackrel{d}{\rightarrow} N(0, \Sigma_{\beta}),
	\end{equation*}
	where $ \Sigma_{\beta}  := \E[ \psi_{\beta}(A; \theta_{0}, q_{\tau})  \psi_{\beta}(A; \theta_{0}, q_{\tau})' ]$. 
\end{lemma}

Lemma \ref{lemma.beta} shows that the parameters of $ F_{Y_{s}|X_{s}Z R = 1} $ are consistently estimated by $ \widehat{\beta} $. Furthermore, it admits an asymptotic linear representation with influence function given by $ \psi_{\beta}(A; \theta_{0}, q_{\tau}) $, which plays a key role in establishing the large sample properties of UQE. Towards this ends, we need the following set of assumptions.

\begin{assumption}\label{assmp.uqe}\ 
	\begin{enumerate}[label=(\alph*)]
		
		\item (i) $ F_{Y|R = 1}(\cdot) $ is absolutely continuous and differentiable over $ y \in \Y $;
		(ii) $ f_{Y|R=1}(\cdot) $ is uniformly continuous;
		(iii) the density $ f_{Y|R=1}(y) $ is strictly bounded away from 0,  three times continuously differentiable in $y$ with uniformly bounded derivatives for  $ y $ in $ \Y_{0}  $, such that $ q_{\tau} \in \Y_{0} $. 
		\item The kernel function $ K_{y}(\cdot) $ is symmetric, continuous, bounded, with a compact support, and such that
		(i)	$ \int K_{y}(y)dy = 1 $; (ii) $ \int y K_{y}(y)dy = 0 $. 
		\item  $ b_{y} \to 0,   \  \log(n) n^{-1} b_{y}^{-1}  \to 0  $, and $  n  b_{y}^{5} \to c_{6} < \infty  $.  
		\item  (i) $ \X $ is compact; (ii) $ G $ is continuously differentiable on $ \X $ with strictly positive density.
	\end{enumerate} 
\end{assumption}
Assumption \ref{assmp.uqe}(a) strengthens Assumption \ref{assmp.ident.hd} and imposes stronger smoothness conditions on the distribution of $ Y_{s} $. 
Assumption \ref{assmp.uqe}(b) states several regularity conditions on kernel functions, which is standard in the literature. 
Assumption \ref{assmp.uqe}(c) specifies admissible rate for the bandwidth parameter. We can choose $ b_{y}= O( n_{s}^{-\kappa})  $, for $ \kappa \in  [1/5, 1/2) $. Assumption \ref{assmp.uqe}(d) imposes support and smoothness conditions for the counterfactual target covariate.

Asymptotic properties of $ \widehat{UQE}$ are formally characterized in the next theorem.  

\begin{theorem} \label{thm.asy.nor}
	Under Assumptions \ref{assmp.data.struct}--\ref{assmp.ident.hd}, \ref{assmp.cr.beta}, and \ref{assmp.uqe}, (i)  the following linear expansions hold, 
	\[  \widehat{UQE}_{q}(\tau, G) - UQE_{q}(\tau, G) =  \dfrac{1}{n} \sum_{i  = 1}^{n} \psi_{q} + B_{q}(\tau, G, b_{y})  + o_{p}(n^{-1/2}).   \]
	Suppose in addition that Assumption \ref{assmp.uqe.2} holds, (ii) then we have 
	\[  \widehat{UQE}_{p}(\tau, G) - UQE_{p}(\tau, G) =  \dfrac{1}{n} \sum_{i  = 1}^{n} \psi_{p} +  B_{p}(\tau, G, b_{y})  + o_{p}(n^{-1/2}),  \]
	where,  $ \psi_{j} $, $j = p,q,$ is defined in Appendix \ref{appn,var.inf},  $ B_{j}(q_{\tau}, G, b_{y}) $ $ := \frac{b_{y}^{2} f_{Y|R}^{''}(q_{\tau}| 1) d_{j}(\theta_{0}, G)  }{2f_{Y|R}^{2}(q_{\tau}|1)}  $ $ \cdot \int y^{2} K_{y}(y) dy$,  and 
	$ d_{j}(\theta_{0}, G)  := \frac{1}{1-Q_{0}}\mathbb{E}[(1-R) \ell(Z) \Lambda_{x}(X, Z_{1}; \beta_{0}) g_{j}(X) ]$, for  $ j = p, q. $
	
	(iii) Therefore, 
	\begin{equation*}
	\sqrt{nb_{y}}  (\widehat{UQE}_{j}(\tau, G) - UQE_{j}(\tau, G) - B_{j}(q_{\tau}, G, b_{y})) \stackrel{d}{\to} N(0, \Sigma_{j}),
	\end{equation*}
	where, $ \Sigma_{j}  :=   \frac{  d_{j}^{2}(\theta_{0}, G) }{f_{Y|R}^{3}(q_{\tau}|1) Q_{0}}  \int K_{y}^{2}(y) dy, $ for $ j = p,q. $
\end{theorem}

From the linear expansions in Theorem \ref{thm.asy.nor}, we conclude that $ UQE $ converges at a rate that is slower than root-$ n $.  This result is mainly driven by the nonparametric estimation of the density $ f_{Y|R=1} $, and therefore, the estimator is nonparametric in essence.  Moreover, the asymptotic expansion includes an asymptotic bias term, $ B(\tau, G, b_{y}) $.  If we assume, as in \citet{firpo2009supplement}, $  n b^{5}_{y} \to 0 $ or $ \kappa <1/5 $, the bias vanishes asymptotically.


\begin{remark}
	Estimators for the asymptotic variance of $ UQE_{p}(\tau, G)$ and $  UQE_{q}(\tau, G)$ can be constructed using their empirical counterparts.  Specifically,  let 
	\begin{equation*}
	\widehat{ \Sigma}_{j} :=  \dfrac{ \widehat{d}_{j,n}(\widehat{ \theta}, G)^{2} }{ \widehat{f}^{3}_{Y|R}(\widehat{q}_{\tau}|1) \E_{n}[R]  }  \int K_{y}^{2}(y) dy,
	\end{equation*}
	where  $\widehat{d}_{j,n}(\widehat{ \theta}, G)  := \mathbb{E}_{n_{a}}[\widehat{\ell}(Z)\Lambda_{x}(W;\widehat{\beta}) \widehat{g}_{j}(X)] $. 
	Under a suitable rate condition on $ b_{y} $, consistency of $ \widehat{ \Sigma}_{j}  $ follows directly from the first two parts of Theorem \ref{thm.asy.nor}.  To achieve better finite-sample performance, we can add the root-n terms of the influence functions to the variance estimator, based on which, we propose the following improved variance estimator, 
	\begin{equation}
	\widehat{\Sigma}_{j, imp}:=  b_{y} \E_{n}[ \widehat{\psi}_{j}(A; \widehat{ \theta }, \widehat{q}_{\tau}, b_{y} )^{2}].
	\end{equation}
	In the above definition, $ \widehat{\psi}_{j}(a; \widehat{ \theta }, \widehat{q}_{\tau}, b_{y} ) $ is a plug-in estimator of the influence function, $  \psi_{j}(A; \theta_{0}, q_{\tau},b_{y} ) $, for $ j = p, q $. 
	A detailed description of the construction of $ \widehat{\psi} $ can be found in the Supplementary Appendix C.   As an alternative, we may conduct inference using a bootstrap procedure, e.g. exchangeable bootstrap. It can be shown that bootstrap approximates not only the leading term but also the higher-order terms in the asymptotic linear representation, which would immediately lead to an improvement over the simple plug-in estimator.  
	
\end{remark}

\begin{remark}
	
	Theorem \ref{thm.asy.nor} implies that tests of the unconditional quantile effect converges at a non-parametric rate in general.  Nonetheless, for the null of zero, positive, and negative effects, we can still construct tests that have power against departures of the null at the parametric rate.  For example, to test the null: $ H_{0} :  UQE_{j}(\tau, G)   = 0  $, it is equivalent to test $  H'_{0} : d_{j}(\beta_{0}, G)   = 0 $, as $UQE_{j}(\tau, G)   = 0  \Leftrightarrow    d_{j}(\beta_{0}, G)   = 0 $, for $ j = p,q $.  From Theorem \ref{thm.asy.nor}, we know that $ \widehat{d}_{j,n}(\widehat{\beta}, G) $ converges at the parametric rate. Moreover, we have
	\begin{equation} \label{eq.test}
	\sqrt{n} \widehat{V}_{d,j}^{-1/2}(\widehat{d}_{j, n}(\widehat{\theta}, G)  - d_{j}(\theta_{0}, G) )  \stackrel{d}{\to } N(0, 1),
	\end{equation}  
	where  $ \widehat{V}_{d,j}  $ is an estimator of  $ V_{d,j} := \E[ \psi_{d,j}(A; \theta_{0}, q_{\tau})^{2}], $ with $ \psi_{d,j} $, $ j = p,q $, defined in  Appendix \ref{appn,var.inf}. The result in \eqref{eq.test}  can be used to test $ H'_{0}  $, applying standard testing procedures. 
\end{remark}

\section{Empirical Application} \label{application}

We apply our identification and estimation methods to a variant of the Mincer's regression.  Our main goal here is to demonstrate the bias from using potential instead of actual labor experience in human capital earnings models. 

Identifying the causal relationship between earnings and human capital accumulation 
has been a focus of labor economic studies for decades.  Traditionally, Mincer's regression has been widely used to quantify the link between labor wage, education and labor market experience. 

Most datasets do not provide respondents' actual work histories.  Therefore, many researchers choose to proxy the variable with potential work experience.  The potential experience measure is usually calculated by subtracting years of schooling plus some constant (typically 6 years) from age. 
Despite the popularity of this practice, many labor economists believe that the return to actual experience tends to be biased when we employ the potential experience as proxy; see e.g.  \citet{regan2009work}.
One of their main argument is that any lapse in labor force participation would be implicitly assumed away when potential experience instead of the actual one is used. There is little reason to believe that the return to employed experience is the same as that of the unemployed period. Hence, it is still preferable to use the actual labor experience.

We use the 1970 wave of IPUMS as our main sample. The data is a 1-in-10,000 national random sample of the population.  The outcome of interest is the natural log of yearly earnings. The target covariate, actual work experience, is missing from IPUMS.  To apply the procedure described in Section \ref{estimation},  we need a dataset where the actual work experience is available. For that purpose, we use the 1972 wave of PSID as cleaned by \citet{hirukawa2020yet}.  Detailed work histories are available in PSID. Therefore, it allows us to recover the actual labor market experience.  However, running analysis directly with PSID may not be ideal due to the fact that it is not nationally representative. Our method is able to address this issue by combining information from both samples.

To estimate $ F_{Y_{s}|X_{s}Z_{1} R = 1} $, we consider the following  specification,
\begin{align}
\PP(log(Income) \leq y ) = \Lambda(\beta_{0} + \beta_{1} & educ + \beta_{2} black + \beta_{3} south \nonumber \\
& + \beta_{4} married + \beta_{5} exper_r + \beta_{6} exper_r^{2} ) \label{application.fyz}
\end{align}
where  $ exper_r $ stands for individual's actual or realized work experience,  $ educ $ denotes the highest grade completed by the respondent, $ black, married, $ and $ south $ are dummy variables which take one if the person is black,  married, and lives in the south, respectively.   

\begin{table}[htbp]
	\caption{Summary Statistics}
	\label{tab: sum.stats}%
	\centering
	\begin{adjustbox}{scale = 0.9,center}
		\begin{tabular}{cccccccc}
			\toprule\toprule
			Variable & \multicolumn{1}{c}{Mean} & \multicolumn{1}{c}{St. Dev.} & \multicolumn{1}{c}{Min} & \multicolumn{1}{c}{25th Pctl.} & \multicolumn{1}{c}{Median} & \multicolumn{1}{c}{75th Pctl.} & \multicolumn{1}{c}{Max} \\
			\midrule
			\multicolumn{8}{c}{Data Source A: The IPUMS Sample} \\
			\midrule
			Income & 8,856.48 & 5,846.45 & 50    & 5,550 & 8,050 & 10,650 & 50,000 \\
			Log(Income) & 8.88  & 0.73  & 3.91  & 8.62  & 8.99  & 9.27  & 10.82 \\
			Age   & 35.6  & 8.3   & 23    & 28    & 35    & 43    & 50 \\
			Education & 11.74 & 2.72  & 5     & 10    & 12    & 13    & 17 \\
			Black & 0.07  & 0.26  & 0     & 0     & 0     & 0     & 1 \\
			South & 0.26  & 0.44  & 0     & 0     & 0     & 1     & 1 \\
			Married & 0.84  & 0.37  & 0     & 1     & 1     & 1     & 1 \\
			Potential Experience & 17.86 & 9.17  & 0     & 10    & 17    & 26    & 39 \\
			\midrule
			\multicolumn{8}{c}{Data Source B: The PSID Sample} \\
			\midrule
			Income & 9,415.42 & 5,620.45 & 50    & 6,000 & 8,598 & 11,721 & 70,000 \\
			Log(Income) & 8.98  & 0.66  & 3.91  & 8.7   & 9.06  & 9.37  & 11.16 \\
			Age   & 34.87 & 8.41  & 23    & 27    & 34    & 42    & 50 \\
			Education & 12.53 & 2.93  & 5     & 11    & 12    & 16    & 17 \\
			Black & 0.26  & 0.44  & 0     & 0     & 0     & 1     & 1 \\
			South & 0.4   & 0.49  & 0     & 0     & 0     & 1     & 1 \\
			Married & 0.9   & 0.29  & 0     & 1     & 1     & 1     & 1 \\
			Potential Experience & 16.34 & 9.49  & 0     & 8     & 16    & 24    & 39 \\
			Actual Experience & 15.98 & 8.39  & 1     & 8     & 16    & 23    & 35 \\
			\bottomrule
			\multicolumn{8}{p{1.1\textwidth}}{\small \textit{Notes}: Summary statistics for IPUMS and PSID. The top panel uses male subsample (aged between 23 to 50) from the 1970 wave of IPUMS with a sample size of 3,504. The bottom panel uses the male subsample (aged between 23 to 50) from the 1972 wave of PSID with a sample size of 1,697.}
		\end{tabular}%
	\end{adjustbox}
\end{table}%

The actual work experience serves as our $ X_{s}$. It enters \eqref{application.fyz} with linear and quadratic terms.  We let  $ (educ, black, south, married) $ be the set of included instruments $ Z_{1} $, and the potential experience, $ exper_p $, be the excluded instrument $ Z_{2}  $.  We provide estimation results when $ \Lambda(\cdot) $ takes either the logistic link or the probit link. To implement the AST estimator, we choose $ j(Z)  = (Z_{1}', Z_{2}')' $, for $ j = k, t, e $.  The density of $ Y_{s} $ is estimated using $ b_{y} = n_{s}^{-0.01} b_{n,0} $, where $ b_{n,0} := 1.06 \min\{\sigma(Y_{s}), interquartile(Y_{s})\} n_{1}^{-0.2} $ is the usual ``rule-of-thumb'' bandwidth.

\begin{figure}[hbt!]
	\centering
	\includegraphics[width=0.85\linewidth]{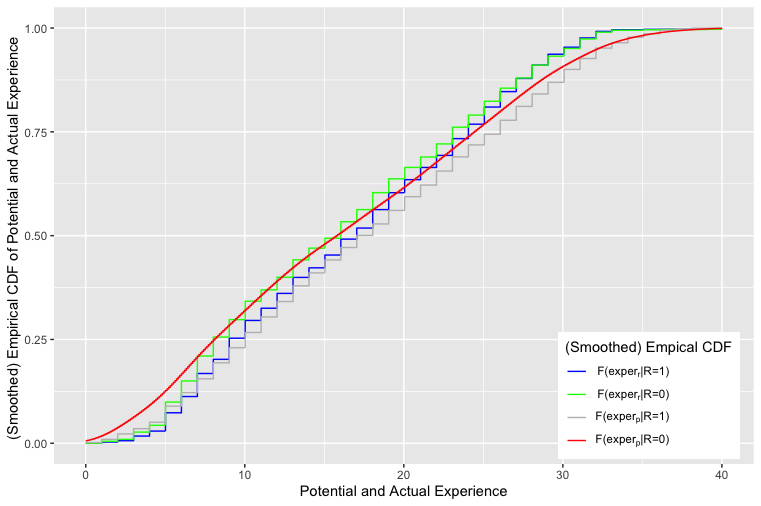}
	\caption{Distributions of  Potential and Actual Experience. }
	\floatfoot{\textit{Notes}: The (smoothed) empirical distributions of potential and actual labor market experience.   The red line depicts the smoothed ECDF of potential experience in the PSID sample (target counterfactual distribution). The green line depicts the ECDF of actual experience in the PSID sample.  The blue line depicts the estimated CDF of actual experience in the IPUMS sample. The grey line depicts the ECDF of potential experience in the IPUMS sample.}
	
	\label{fig:ecdf}
\end{figure}

\begin{table}[htbp]
	\centering
	\caption{Estimation Results}
	\label{tab: est.res}%
	\begin{tabular}{lccccccc}
		\toprule\toprule
		\multicolumn{1}{l}{Quantile Level} & \multicolumn{1}{c}{0.25} & \multicolumn{1}{c}{0.5} & \multicolumn{1}{c}{0.75} &       & \multicolumn{1}{c}{0.25} & \multicolumn{1}{c}{0.5} & \multicolumn{1}{c}{0.75} \\
		\midrule
		& \multicolumn{3}{c}{Logit Link } &       & \multicolumn{3}{c}{Probit Link} \\
		\hline
		MDS   &       &       &       &       &       &       &  \\
		\hline
		$UQE_{2s}(\tau)$ & -0.0490 & -0.0303 & -0.0153 &       & -0.0606 & -0.0308 & -0.0157 \\
		& (0.0244) & (0.0116) & (0.0071) &       & (0.0277) & (0.0119) & (0.0071) \\
		$ H_0: UQE $ = 0 & 0.0426 & 0.0080 & 0.0292 &       & 0.0272 & 0.0087 & 0.0260 \\
		\hline
		MQS   & \multicolumn{3}{c}{}  &       & \multicolumn{3}{c}{} \\
		\hline
		$UQE_{2s}(\tau)$ & -0.0458 & -0.0297 & -0.0149 &       & -0.0575 & -0.0300 & -0.0152 \\
		& (0.0112) & (0.0065) & (0.0056) &       & (0.0128) & (0.0065) & (0.0056) \\
		$ H_0: UQE $ = 0 & 0.0000 & 0.0000 & 0.0075 &       & 0.0000 & 0.0000 & 0.0064 \\
		\hline
		MLS   & \multicolumn{3}{c}{}  &       & \multicolumn{3}{c}{} \\
		\hline
		$UQE_{2s}(\tau)$ & 0.0240 & 0.0189 & 0.0150 &       & 0.0268 & 0.0191 & 0.0154 \\
		& (0.0033) & (0.0019) & (0.0017) &       & (0.0035) & (0.0019) & (0.0016) \\
		$ H_0: UQE $ = 0 & 0.0000 & 0.0000 & 0.0000 &       & 0.0000 & 0.0000 & 0.0000 \\
		\bottomrule
		 \multicolumn{8}{p{0.9\textwidth}}{\small \textit{Notes}: In each panel, the first two rows report point estimates and standard error using our two sample estimator.  The last row of each panel reports the $ p $-value associated with the Wald test of zero effect.}
	\end{tabular}%
\end{table}%

\begin{figure}[!ht]
	\centering
	
	\begin{subfigure}{0.5\textwidth}
		\centering
		\includegraphics[width=1\linewidth]{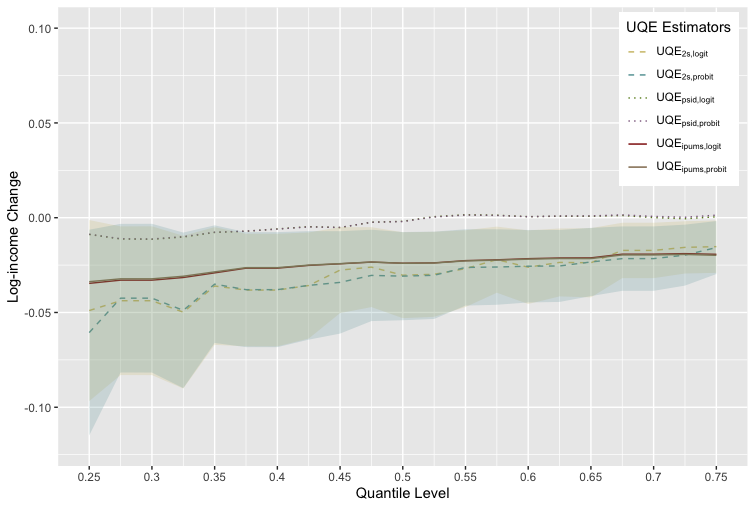}
		\caption{MDS}
		\label{fig:sub1}
	\end{subfigure}%
	\begin{subfigure}{0.5\textwidth}
		\centering
		\includegraphics[width=1\linewidth]{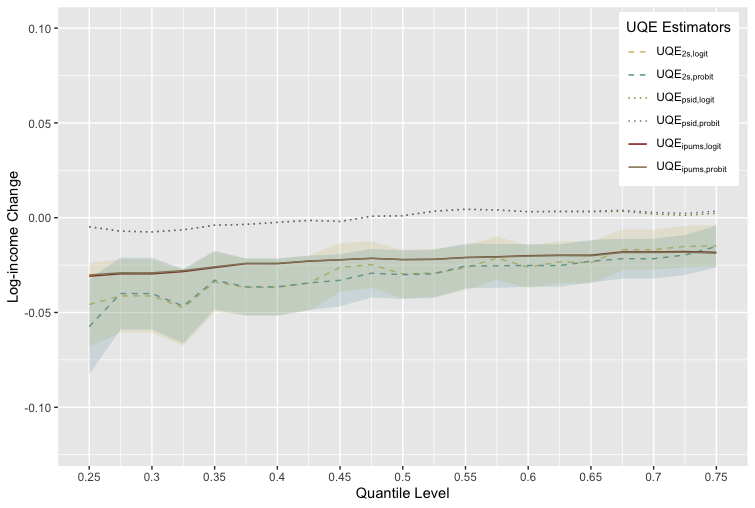}
		\caption{MQS}
		\label{fig:sub2}
	\end{subfigure}
	\\
	\begin{subfigure}{0.70\textwidth}
		\centering
		\includegraphics[width=1 \linewidth]{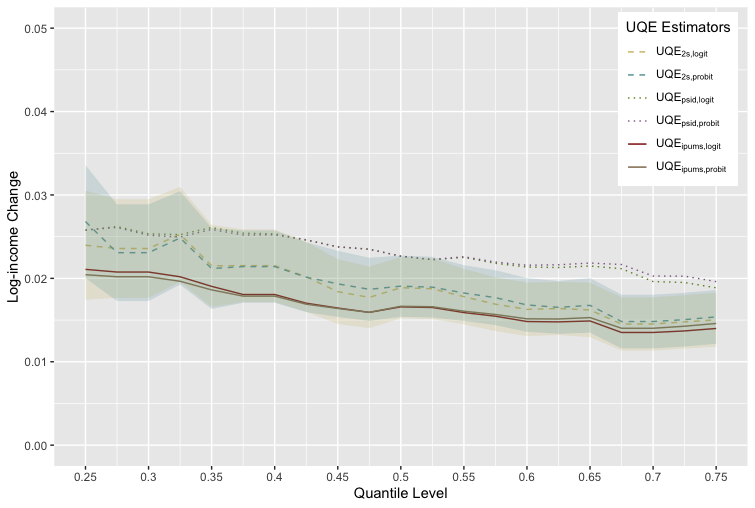}
		\caption{MLS}
		\label{fig:sub3}
	\end{subfigure}
	\caption{Unconditional Quantile Effect of Actual Experience on Log(Earnings). } 
	\label{fig:uqe.plots}
	\floatfoot{\textit{Notes}: The UQE of labor market experience on log earnings.  The top left panel: results for UQE with MDS. The top right panel: results for UQE with MQS. The bottom panel: results for UQE with MLS. All three plots contain the UQE of potential experience based on IPUMS (solid lines), the UQE of actual experience based on PSID (dotted lines),  the two-sample UQE (dashed lines), and the two-sided 95\% confidence intervals based on the improved variance estimator (shaded area).}
\end{figure}

Table \ref{tab: sum.stats} reports the descriptive statistics for the two samples. We use only the data on men aged between 23 and 50 when the surveys are taken, to ensure that common support condition holds. This leaves us with a sample of $ n_{s} =  3,504$ respondents for IPUMS and $ n_{a} = 1,697 $ for PSID.
There are considerable differences between the two datasets. Individuals who are black, married and/or lives in the south are over-represented in PSID compared to the nationally representative IPUMS.  On average, an individual in PSID has 0.36 years more potential experience than actual experience.  

For UQE with MDS and MQS,  we choose the smoothed empirical distribution of $ exper_{p} $ in the PSID sample (assuming it is fixed and known) as the target counterfactual distribution, i.e. $ G = F_{Z_{2}|R = 0} $. As depicted in Figure \ref{fig:ecdf}, we find that less-experienced workers  tend to have even fewer years of experience in the counterfactual scenario than in the status quo, and the opposite is true for workers closer to the right tail of the distribution.

We report estimation results in Table \ref{tab: est.res} and Figure \ref{fig:uqe.plots}.  A few remarks are in order. First, our estimates suggest that the counterfactual effect of a marginal shift in the distribution of actual experience is heterogeneous across income groups. The effect is larger in magnitude for the lower-income groups as expected.    When MDS and MQS are considered, the quantile effects are uniformly negative and the shapes of the effect curves are similar. The marginal shift could decrease the (log) earning by anything between 0.015 and 0.061 across income quantiles. For reference, when  logistic link is assumed, the marginal effect of MDS amounts to a  reduction of 4.8\% in annual earnings for individuals the first quartile,  3.0\% at the median, and  1.5\%  at the third quartile, respectively. For MLS, two-sample estimates are bounded between the two sets of one-sample estimates. The marginal upward shift in the actual experience would increase a median worker's income by 1.9\%.

Next, we consider the bias caused by using potential experience in lieu of the actual experience. We note that, one-sample UQE estimates based on IPUMS tend to be smaller in magnitude at lower income quantiles than that based on the combined data. Eventually, the two estimators converge at higher income levels.  In the context of our application, there does not seem to be sufficient evidence to support the claim that the bias is statistically significant.\footnote{Our analysis is local to the direction of counterfactual change, and therefore, does not allow the insignificance result to be extrapolated globally. For the same reason, the comparison between $ UQE_{2s} $ and $ UQE_{psid} $ is not meaningful.}

\section{Concluding Remarks} \label{conclusion}
In this paper, we propose a framework to identify and estimate unconditional quantile policy effect under data combination.  We establish the identification of UQE under two main conditions: a rank similarity assumption and a conditional independence assumption, based on which,  we provide estimators for the identified UQE and derive their large sample properties. 

Our current approach can be extended in the following directions. First, although we have restricted our attention to the quantile effect throughout this paper, our results can be easily extended to other statistical functionals such as mean, interquartile, and inequality measures.  It would be interesting to see how the identification requirements change with respect to the functional we adopt. 
Second, we have focused exclusively on the pointwise identification and inference.  While extension to uniform results seem straightforward, it comes at a cost of stronger cross-sample restrictions.  Under such assumptions, conditional quantile regression is likely feasible. Comparing conditional and unconditional quantile effects,  as in \citet{firpo2009unconditional}, under our two-sample structure, would also be an interesting direction for future research. 

\appendix
	\pagebreak

		\part*{\LARGE Appendix} 
	\section{Proofs of Lemmas and Theorems in Section \ref{identification}}

	\textit{Proof of Lemma \ref{lemma.ident.fyw}: }
		We provide proof for the nonparametric identification here.  The proof for the parametric case follows along exactly the same line and is omitted. We shall show \eqref{eq.ident.fyw} first, 
		\begin{align*}
		\mathbb{E}[ 1(Y \leq q_{\tau})  |Z, R= 1] &  = \mathbb{E}[ \mathbb{E}[ 1(Y_{s} \leq q_{\tau})  |X_{s}, Z, R= 1] | Z, R = 1] \\
		& = \mathbb{E}[ \mathbb{E}[ 1(Y_{s} \leq q_{\tau})  |X_{s}, Z_{1}, R= 1]| Z, R = 1]  \\
		& = \mathbb{E}[ \Lambda(X_{s}, Z_{1})  | Z, R = 1]  \\
		& = \mathbb{E}[ \Lambda(X_{a}, Z_{1})  | Z, R = 0]  \\
		& = \mathbb{E}[ \Lambda(W) | Z, R = 0],
		\end{align*}
		where the second equality is by Assumption \ref{assmp.ident.fyw.1},  and the fourth line follows by Assumption \ref{assmp.data.struct}(e). Likewise for \eqref{eq.ident.fyw.1}, 
		\begin{align*}
		\mathbb{E}[ R1(Y \leq q_{\tau}) |Z ]  & = \mathbb{E}[ 1(Y \leq q_{\tau})  |Z, R= 1] \cdot \mathbb{P}[R = 1 |Z] \\
		& = \mathbb{E}[ \Lambda(W) | Z, R = 0] \cdot r(Z) \\
		& = \mathbb{E}[ (1-R) \Lambda(W) | Z] \cdot \dfrac{r(Z)}{1-r(Z)}.
		\end{align*}
		
		Thus, Lemma \ref{lemma.ident.fyw} follows immediately from \eqref{eq.ident.fyw}  (or \eqref{eq.ident.fyw.1}) and Assumption \ref{assmp.ident.fyw.2}.   $ \blacksquare $ \\

	\begin{lemma}\label{lemma.bdd.comp}
	Suppose (i)	$ \Lambda(w; \beta) $ is  measurable with respect to $ w $ for all $ \beta \in \Theta_{\beta} $; (ii)   $ W $ is bounded complete for $ Z $, relative to the auxiliary population; (iii) $ \Lambda(w; \beta)$ is differentiable with respect to $ \beta $; and (iv) $ \partial \Lambda(\cdot; \beta)/ \partial \beta $ is uniformly bounded and $ \partial \Lambda(\cdot; \beta)/ \partial \beta \not\equiv 0 $ for all $ \beta \in \Theta_{\beta} $. Then, under Assumptions \ref{assmp.data.struct}  and  \ref{assmp.ident.fyw.1}, $ \beta_{0} $ can be uniquely identified from \eqref{eq.ident.fyw} or \eqref{eq.ident.fyw.1}. 
    \end{lemma}    

\noindent \textit{Proof of Lemma \ref{lemma.bdd.comp}: }
	From Lemma \ref{lemma.ident.fyw}, we know that $ \beta_{0} $ solves \eqref{eq.ident.fyw} or \eqref{eq.ident.fyw.1}. It remains to show uniqueness. Suppose,  there is $ \beta_{1} $, $ \beta_{1}  \neq \beta_{0}$, that solves \eqref{eq.ident.fyw}, then, $ \E[\Lambda(W; \beta_{1}) - \Lambda(W; \beta_{0}) | Z, R = 0] = 0.  $ By MVT, this and (iii) implies that $  \E[\partial \Lambda(W; \beta) / \partial \beta |_{\beta  = \widetilde{\beta}} | Z, R = 0] (\beta_{1} - \beta_{0}) = 0$, for some value between $ \beta_{0} $ and $ \beta_{1} $. Condition (i), (ii), and (iv) then imply that $ \E[\partial \Lambda(W; \beta) / \partial \beta |_{\beta  = \widetilde{\beta}} | Z, R = 0]  \not\equiv 0 $, which leads to a contradiction.  $ \blacksquare $ \\

\noindent \textit{Proof of Theorem \ref{thm.main}: }
    We shall first prove the identification result for a fixed counterfactual distribution. Next, we take the derivative of the counterfactual experiments with respect to $ t $. The result of Theorem \ref{thm.main} then follows by the fact that Hadamard derivative operator of the quantile functional is linear.  For any $ t \leq t_{0} $, fix  $ \epsilon_{t}  \in \Phi^{\ast}$, and we have that
	\begin{align*}
	F&_{\widetilde{Y}_{s,t}|R}(q_{\tau}|1) \\
	&= \int P(g_{s}( \widetilde{X}_{s,t}, \widetilde{Z}_{1,t}, \widetilde{\epsilon}_{s,t}) \leq q_{\tau} | \widetilde{X}_{s,t} = x , \widetilde{Z}_{1,t} = z_{1}, \widetilde{R}_{t} =1  ) d F_{\widetilde{X}_{s,t} \widetilde{Z}_{1,t}|\widetilde{R}_{t}}(x, z_{1}|1) \\
	&= \int P(g_{s}(G^{-1}_{t}(\widetilde{U}_{s,t}), \widetilde{Z}_{1,t}, \widetilde{\epsilon}_{s,t}) \leq q_{\tau} | \widetilde{U}_{s,t} = u , \widetilde{Z}_{1,t} = z_{1}, \widetilde{R}_{t} =1  ) d F_{\widetilde{U}_{s,t} \widetilde{Z}_{1,t}|\widetilde{R}_{t}}(u, z_{1}|1) \\
	& = \int P(g_{s}(G^{-1}_{t}(u), Z_{1}, \epsilon_{s}) \leq q_{\tau} |  Z_{1} = z_{1}, R = 1 ) d F_{U_{s} Z_{1}|R }(u, z_{1}|1)\\
	& = \int P(g_{s}(G^{-1}_{t}(u), Z_{1}, \epsilon_{s}) \leq q_{\tau} |  U_{s}  = u, Z_{1} = z_{1}, R = 1 ) d F_{U_{s} Z_{1}|R }(u, z_{1}|1)\\
	& = \int P(g_{s}(X_{s}, Z_{1}, \epsilon_{s}) \leq q_{\tau} | X_{s} = G^{-1}_{t}(u),  Z_{1} = z_{1}, R = 1 ) d F_{U_{s}Z_{1}|R}(u, z_{1}|1)\\
	& = \int P(g_{s}(X_{s}, Z_{1}, \epsilon_{s}) \leq q_{\tau} | X_{s} = G^{-1}_{t}(u),  Z_{1} = z_{1}, R = 1 ) d F_{U_{s}Z|R}(u, z|1)\\
	& = \int P(g_{s}(X_{s}, Z_{1}, \epsilon_{s}) \leq q_{\tau} | X_{s} = G^{-1}_{t}(F_{X_{s}}(x)),  Z_{1} = z_{1}, R = 1 )  d F_{X_{s}Z|R}(x, z|1)\\
	& = \int F_{Y_{s}|X_{s}Z_{1}R}(q_{\tau}| G^{-1}_{t}(F_{X_{s}|R}(x|1)), Z_{1}, 1)  d F_{X_{s}Z|R}(x, z|1)\\
	& = \int F_{Y_{s}|X_{s}Z_{1}R}(q_{\tau}| G^{-1}_{t}(F_{X_{s}|R}(x|1)), Z_{1}, 1) \dfrac{r(z)(1-Q_{0})}{Q_{0}(1-r(z))}d F_{XZ|R}(x, z|0)\\
	& = \mathbb{E}\left[ F_{Y_{s}|X_{s}Z_{1}R}(q_{\tau}| G^{-1}_{t}(F_{X_{s}|R}(X|1)), Z_{1}, 1) \dfrac{r(Z)(1-Q_{0})}{Q_{0}(1-r(Z))}  | R = 0\right]\\
	& = \dfrac{1}{1-Q_{0}}\mathbb{E}\left[(1-R)\ell(Z)\cdot F_{Y_{s}|X_{s}Z_{1}R}(q_{\tau}| G^{-1}_{t}(F_{X|R}(X|1)), Z_{1}, 1)\right],
	\end{align*} 
	where the second line follows by the definition of $ F_{\widetilde{Y}_{s,t}}(q_{\tau}) $, the third one comes from the definition of $ \widetilde{U}_{s,t} $  and a change of variable from $ x $ to $  u $, the fourth equality follows by the construction of $ \Phi^{\ast} $ and Assumptions \ref{assmp.ident.ex.cont}(a) and (b), the fifth line is again by Assumption \ref{assmp.ident.ex.cont}(a), the eighth one follows by the definition of $ U_{s} $ and standard change-of-variable argument,  the tenth line is by  Assumptions \ref{assmp.data.struct}(a)--(c) and Bayes' Law.

	To obtain the marginal distributional effect, we take derivative of $ F_{\widetilde{Y}_{s}|R}^{G_{t}}(q_{\tau}|1) $ with respect to $ t $ and evaluate it at $ t=0 $. For the marginal distributional shift, 
	\begin{align*}
	\left.	\dfrac{\partial F_{\widetilde{Y}_{s,t}|R}(q_{\tau}|1)}{\partial t }\right\vert_{t= 0} = & \int \left. \dfrac{\partial F_{Y_{s}|X_{s}Z_{1}R}(q_{\tau}| x, z_{1}, 1) }{\partial x} \cdot \dfrac{\partial G^{-1}_{t,p}( F_{X_{s}|R}(x|1) )}{\partial t}\right\vert_{t=0} \\
	&\qquad \qquad \qquad \qquad \qquad \qquad \qquad \cdot \dfrac{r(z)(1-Q_{0})}{Q_{0}(1-r(z))} d F_{W|R}(w|0) \\
	=  & \dfrac{1}{1-Q_{0}}\mathbb{E}\left[(1-R)\ell(Z)\cdot \dfrac{\partial \Lambda( X, Z_{1})}{\partial x }  \cdot \dfrac{\partial G^{-1}_{t,p}( F_{X_{s}|R}(x|1) )}{\partial t}|_{t=0}    \right].
	\end{align*}
	
	Observe that $  \frac{\partial G^{-1}_{t,p}(\cdot )}{\partial t}|_{t=0} $ is the pathwise derivative of  the inverse map $ H \mapsto H^{-1} $ at $ F_{X_{s}|R = 1} $ in the direction of $ G(\cdot) - F_{X_{s}|R = 1} $. By Lemma 3.9.23 in \citet{van1996weak}, the inverse map is Hadamard differentiable under the conditions specified in the theorem, with the derivative map given by, 
	\begin{equation*}
	\phi  \mapsto - (\phi /h) \circ H^{-1},
	\end{equation*}
	where $ h $ is the first-order derivative of $ H $.  Let $ 	\phi(\cdot) = G(\cdot) - F_{X_{s}|R = 1}(\cdot)   $ and $ H = F_{X_{s}|R = 1} $, it follows immediately that,  for all $ u \in [0,1] $,
	\begin{equation*}
	\left.	\frac{\partial G^{-1}_{t,p}(u )}{\partial t}\right\vert_{t=0}  = - \dfrac{G( F_{X_{s}|R = 1}^{-1}(u) ) - u   ) }{f_{X_{s}|R = 1}(  F_{X_{s}|R = 1}^{-1}(u ) )},
	\end{equation*}
	and hence, for all $ x \in \X $,
	\begin{equation*}
	\left.	\dfrac{\partial G^{-1}_{t,p}( F_{X_{s}|R}(x|1) )}{\partial t}\right\vert_{t=0}  =  \dfrac{ F_{X_{s}|R}(x|1) -  G( x ) }{f_{X_{s}|R = 1}( x )}.
	\end{equation*}
	
	Analogously,  for marginal quantile shift, 
	\begin{align*}
	\left.	\dfrac{\partial F_{\widetilde{Y}_{s,t}|R}(q_{\tau}|1)}{\partial t }\right\vert_{t= 0}  = & \int \dfrac{\partial F_{Y_{s}|X_{s}Z_{1}R}(q_{\tau}| x, z_{1}, 1) }{\partial x} \cdot \left. \dfrac{\partial G^{-1}_{t,q}( F_{X_{s}|R}(x|1) )}{\partial t}\right\vert_{t=0} \\
	&\qquad \qquad \qquad \quad \qquad \qquad \qquad \qquad \cdot\dfrac{r(z)(1-Q_{0})}{Q_{0}(1-r(z))} d F_{W|R}(w|0) \\
	 = & \dfrac{1}{1-Q_{0}}\mathbb{E}\left[(1-R)\ell(Z) \cdot \dfrac{\partial \Lambda(X, Z_{1})}{\partial x } \cdot (G^{-1}(F_{X|R =1}(X))  - X) \right],
	\end{align*}	
	where the second equality follows from Lemma \ref{lemma.ident.fyw} and the definition of $  G^{-1}_{t,q}(\cdot) $. 
	
	To identify $ F_{X_{s}|R =1} $, we exploit the following fact
	\begin{align*}
	F_{X_{s}|R =1}(\cdot)
	& = \mathbb{E}[  1(X\leq \cdot)   |R = 1]  \\
	& = \int_{\Z} \int_{\X}  1(x\leq \cdot)   dF_{X|Z R}(x|z,1) d F_{Z|R}(z|1) \\
	& = \int_{\Z}  \int_{\X}   1(x\leq \cdot)  \cdot \dfrac{(1-Q_{0})r(z)}{Q_{0}(1-r(z))} dF_{X|Z R}(x|z,0) d F_{Z|R}(z|0) \\
	& = \mathbb{E}\left[  \dfrac{(1-Q_{0})r(Z)}{Q_{0}(1-r(Z))} 1(X\leq \cdot)   |R =0\right]  \\
	& = \dfrac{1}{1-Q_{0}}\mathbb{E}\left[(1-R)\ell(Z)1(X\leq \cdot) \right],
	\end{align*}
	where the third line is due to Assumption \ref{assmp.data.struct} and Bayes' Law. 
	
	Theorem \ref{thm.main} then follows from Assumption \ref{assmp.ident.hd}, $ q_{\tau} =  F_{Y_{s}|R=1}^{-1}(\tau) $, and  the fact that the Hadamard derivative of the quantile functional is $ \nu_{\tau}'(\phi )  = - \dfrac{ \phi }{f_{Y_{s}|R=1}} \circ F_{Y_{s}|R=1}^{-1}(\tau)$, which is linear in $ \phi. $ $ \blacksquare $  \newline

\noindent \textit{Proof of Theorem \ref{thm.main.discrete}: }
	First, we fix $  U_{s} \in \mathcal{U}_{s}  $ and   $ \phi_{t}  \in \Phi^{\ast}$, for $ t \leq t_{0} $.  By construction, there exists $ \widetilde{U}_{s,t} \in \mathcal{\widetilde{U}}_{s,t}  $ such that $ ( \widetilde{U}_{s,t} | Z_{1}, R = 1) \stackrel{d}{=} (U_{s}| Z_{1}, R = 1 ) $. Now we rewrite $ F_{\widetilde{Y}_{s,t}|R = 1} \in \F_{\widetilde{Y}_{s,t}|R= 1} $ in terms of $ U_{s} $ and $ Z_{1} $. Let  $ x^{0} = -\infty $, and we have that
	\begin{align}
	F&_{\widetilde{Y}_{s,t}|R= 1}(q_{\tau})  \nonumber \\
	=&   \int P(g_{s}(\widetilde{X}_{s,t}, \widetilde{Z}_{1,t}, \widetilde{\epsilon}_{s,t}) \leq q_{\tau} |\widetilde{X}_{t} = x, \widetilde{Z}_{1,t} = z_{1}, \widetilde{R}_{t} = 1) d F_{\widetilde{X}_{s,t}\widetilde{Z}_{1,t}|\widetilde{R}_{t}} (x, z_{1}|1) \nonumber\\
	=&   \sum_{j = 1}^{l}\int P(g_{s}(\widetilde{X}_{s,t}, \widetilde{Z}_{1,t}, \widetilde{\epsilon}_{s,t}) \leq q_{\tau} |\widetilde{X}_{s,t} = x^{j}, \widetilde{Z}_{1,t} = z_{1}, \widetilde{R}_{t} = 1)\nonumber \\
	&  \qquad   \cdot  P(\widetilde{U}_{s,t} \in ( G_{t,p}(x^{j-1}),   G_{t,p}(x^{j})] |\widetilde{Z}_{1,t} = z_{1}, \widetilde{R}_{t} = 1)d F_{\widetilde{Z}_{1,t}|\widetilde{R}_{t}}(z_{1}|1) \nonumber\\
	=&  \sum_{j = 1}^{l} \int P(g_{s}( X_{s}, Z_{1}, \epsilon_{s}) \leq q_{\tau} | X= x^{j}, Z_{1} = z_{1}, R= 1) \nonumber\\
	&  \qquad    \cdot P(U_{s} \in ( G_{t,p}(x^{j-1}),   G_{t,p}(x^{j})] |Z_{1} = z_{1}, R = 1)d F_{Z_{1}|R}(z_{1}|1) \nonumber\\
	=&  \sum_{j = 1}^{l} \int F_{Y_{s}|X_{s}Z_{1}R}  (q_{\tau}|x^{j}, z_{1}, 1)\cdot ( P(U_{s} \in ( F_{X_{s}|R}(x^{j-1}|1),  F_{X_{s}|R}(x^{j}|1)] |Z_{1} = z_{1}, R = 1)\nonumber \\
	&   + P(U_{s} \in ( G_{t,p}(x^{j-1}),   G_{t,p}(x^{j})] |Z_{1} = z_{1}, R = 1)\nonumber \\
	&   - P(U_{s} \in ( F_{X_{s}|R}(x^{j-1}|1),  F_{X_{s}|R}(x^{j}|1)] |Z_{1} = z_{1}, R = 1) )d F_{Z_{1}|R}(z_{1}|1)   \nonumber \\
	= & F_{Y_{s}|R}(q_{\tau}|1)  - \sum_{j = 1}^{l}\int \Lambda(x^{j}, z_{1})  (  P(U_{s} \in ( G_{t,p}(x^{j-1}),   G_{t,p}(x^{j})] |Z_{1} = z_{1}, R = 1) \nonumber \\
	&   - P(U_{s} \in ( F_{X_{s}|R}(x^{j-1}|1),  F_{X_{s}|R}(x^{j}|1)] |Z_{1} = z_{1}, R = 1) )d F_{Z_{1}|R}(z_{1}|1).  \label{eq.thm.discrete}
	\end{align}
	For the second term on the right hand side of the last equality, we have
	\begin{align*}
	P(U_{s} & \in ( G_{t,p}(x^{j-1}),   G_{t,p}(x^{j})] \mid Z_{1} = z_{1}, R = 1)   \\
	& - P(U_{s} \in ( F_{X_{s}|R}(x^{j-1}|1),  F_{X_{s}|R}(x^{j}|1)] \mid Z_{1} = z_{1}, R = 1) )  \\
	= &( F_{U_{s}|Z_{1} R}(  G_{t,p}(x^{j}) \mid z_{1}, 1 ) - F_{U_{s}|Z_{1} R}(   F_{X_{s}|R}(x^{j}|1) \mid  z_{1}, 1 ) \\
	& - (  F_{U_{s}|Z_{1} R}(  G_{t,p}(x^{j-1}) \mid z_{1}, 1 ) )  - F_{U_{s}|Z_{1} R}(   F_{X_{s}|R}(x^{j-1}|1) \mid  z_{1}, 1 ) )\\
	= &  ( G_{t,p}(x^{j}) -   F_{X_{s}|R}(x^{j}|1) ) \cdot f_{U_{s}|Z_{1} R}(  \widetilde{u}_{j,t} \mid z_{1}, 1 )  \\
	& -    ( G_{t,p}(x^{j-1}) -   F_{X_{s}|R}(x^{j-1}|1) ) \cdot f_{U_{s}|Z_{1} R}(  \widetilde{u}_{j-1,t} \mid z_{1}, 1 )  \\
	= &  t \cdot ( G(x^{j}) -   F_{X_{s}|R}(x^{j}|1) ) \cdot f_{U_{s}|Z_{1} R}(  \widetilde{u}_{j,t} \mid z_{1}, 1 )  \\
	& -    t \cdot ( G(x^{j-1}) -   F_{X_{s}|R}(x^{j-1}|1) ) \cdot f_{U_{s}|Z_{1} R}(  \widetilde{u}_{j-1,t} \mid z_{1}, 1 ),
	\end{align*}
	where $ \widetilde{u}_{j,t} $ is some value between $  G_{t,p}(x^{j})   $ and $ F_{X_{s}|R}(x^{j}|1) $, and is potentially dependent on $ z_{1} $. The last equality is due to MVT.  Using the above result,  \eqref{eq.thm.discrete}  becomes 
	\begin{align*}
	F_{Y_{s}|R}(q_{\tau}|1)  & - \sum_{j = 1}^{l} t \cdot \int \Lambda(x^{j}, z_{1}) \cdot  (   ( G(x^{j}) -   F_{X_{s}|R}(x^{j}|1) ) \cdot f_{U_{s}|Z_{1} R}(  \widetilde{u}_{j, t} \mid z_{1}, 1 )  \\
	& -   ( G(x^{j-1}) -   F_{X_{s}|R}(x^{j-1}|1) ) \cdot f_{U_{s}|Z_{1}R}(  \widetilde{u}_{j-1, t} \mid z_{1}, 1 )    )d  F_{Z_{1}|R}(z_{1}|1) \\
	=  F_{Y_{s}|R}(q_{\tau}|1)  & - \sum_{j = 2}^{l} t \cdot \int (\Lambda(x^{j-1}, z_{1})  - \Lambda(x^{j}, z_{1}) )\\
	& \cdot    ( G(x^{j-1}) -   F_{X_{s}|R}(x^{j-1}|1) ) \cdot f_{U_{s}|Z_{1}R}(  \widetilde{u}_{j-1, t} \mid z_{1},  1 )  d F_{Z_{1}|R}(z_{1}|1),
	\end{align*}
	where the equality follows by rearranging terms, the fact that $ G(x^{0}) = F_{X_{s}|R}(x^{0}|1)  = 0$, and that $ G(x^{l}) = F_{X_{s}|R}(x^{l}|1)  = 1$.  %
	The pathwise derivative can thus be calculated as 
	\begin{align*}
	\lim_{t \downarrow 0} &\dfrac{F_{\widetilde{Y}_{s,t}|R= 1}(q_{\tau})   -  F_{Y_{s}|R}(q_{\tau}|1) }{t}  \nonumber \\
	= & \sum_{j = 2}^{l}  \int (\Lambda(x^{j-1}, z_{1})  - \Lambda(x^{j}, z_{1}) ) \\
	& \qquad \cdot    ( G(x^{j-1}) -   F_{X_{s}|R}(x^{j-1}|1) ) d F_{Z_{1} | U_{s} R}(  z_{1} \mid F_{X_{s}|R}(x^{j-1}|1), 1 ),
	\end{align*}
	where the second line is due to the dominated convergence theorem,  Bayes's Law,  and the fact that $ U_{s}|R = 1 $ follows the standard uniform distribution. 
	Therefore, by Lemma \ref{lemma.ident.fyw} and the linearity of $ \nu_{\tau}'(\cdot) $,
	\begin{multline*}
	UQE_{p}(\tau, G)  \in \left[\inf\limits_{U_{s}\in {\mathcal{U}}_{s}}  \sum_{j = 2}^{l}  \int h_{q_{\tau}}(x^{j}, x^{j-1}, z_{1}) d F_{Z_{1} | U_{s} R}(  z_{1}  \mid F_{X_{s}|R}(x^{j-1}|1), 1 ) , \right. \\
	\left. \sup \limits_{U_{s}\in {\mathcal{U}}_{s}} \sum_{j = 2}^{l}  \int h_{q_{\tau}}(x^{j}, x^{j-1}, z_{1})  d F_{Z_{1} | U_{s} R}(  z_{1}  \mid F_{X_{s}|R}(x^{j-1}|1), 1 )  \right]. 
	\end{multline*}

	Using a similar argument as in the proof of Theorem 5 in \citet{rothe2012partial}, we can show that for $ j = 1,\dots, l $, $ \{ F_{Z_{1}| U_{s}R}(z_{1}| U_{s} =F_{X_{s}|R}(x^{j}|1), R = 1): U_{s} \in \mathcal{U}_{s}\}$ is the set of all multivariate distribution functions with support equal to $ \supp(F_{Z_{1}|R =1}) $.  To see this, note that for $ j = 1,\dots, l $, $  F_{Z_{1}| U_{s}R}(\cdot | U_{s} =F_{X_{s}|R}(x^{j}|1), R = 1)  = C^{U_{s}}_{1}( F_{X_{s}|R}(x^{j}|1), F_{Z_{1}|R}(\cdot|1) )$, where the conditional copula, $  C^{U_{s}} $, is defined by $ C^{U_{s}}(F_{U_{s}|R}(u|1), F_{Z_{1}|R}(z_{1}|1)) $ $:= F_{U_{s}Z_{1}|R}(u, z_{1}|1) $, and $ C^{U_{s}}_{1} $ is the partial derivative of $ C^{U_{s}} $ with respect to the first argument.  By the construction of $ \Phi $, the set of $ C^{U_{s}}(\cdot, \cdot)  $ for $ U_{s} \in  {\mathcal{U}}_{s} $ is equivalent to the identified set of the conditional copula of $ X_{s} $ and $ Z_{1} $ given $ R = 1 $, $ C^{X_{s}}(\cdot, \cdot) $, where $ C^{X_{s}}(F_{X_{s}|R}(x|1), F_{Z_{1}|R}(z_{1}|1)) := F_{X_{s}Z_{1}|R}(x, z | 1)  $, for all $ x \in \{x^{1},\dots, x^{l}  \} $. Then, the desired result follows by applying an extension of Theorem 2.2.7 in \citet{nelsen2007introduction}.
	

	Without loss of generality, we focus on the upper bound for now.   By appropriately choosing  Dirac measures with unit masses on $ \{z^{\ast}_{j}\}_{j\in\J_{+}} $ and $ \{z^{\dagger}_{j}\}_{j\in\J_{-}} $, It is straightforward to show that,
	\begin{align}\label{eq.sup.u} 
	\sup \limits_{U_{s}\in \mathcal{U}_{s}} &\sum_{j = 2}^{l}  \int h_{q_{\tau}}(x^{j}, x^{j-1}, z_{1}) d F_{Z_{1} | U_{s} R}(  z_{1}  \mid F_{X_{s}|R}(x^{j-1}|1), 1 ) \nonumber \\
	&= \sum_{j \in \J_{+}}  h_{q_{\tau}}(x^{j}, x^{j-1}, z^{\ast}_{1,j}) +  \sum_{j \in \J_{-}}  h_{q_{\tau}}(x^{j}, x^{j-1}, z^{\dagger}_{1,j}).
	\end{align}
	The right hand side of \eqref{eq.sup.u} is identified under the support condition in Assumption \ref{assmp.data.struct}(a).
	The proof for the lower bound follows by an analogous argument.  	$ \blacksquare $
	\bigskip
	
	\section{Asymptotic Linear Representation of UQE Estimators} \label{appn,var.inf}
	We specify additional regularity conditions in Theorem \ref{thm.asy.nor} and provide linear expansions for $ \widehat{UQE}_{p} $ and $ \widehat{UQE}_{q} $ in this section, the proofs of which are contained in the Supplementary Appendix.

  	\begin{assumption}\label{assmp.uqe.2}\
  	\begin{enumerate}[label=(\alph*)]
  		\item  $ f_{X|ZR = 0} $ is uniformly bounded,  twice continuously differentiable with uniformly bounded first and second order derivatives on $ \X\Z $.   
  		\item (i) $ K_{x}(\cdot) $ is a second order symmetric kernel function; (ii) the support of $ K_{x}  $ is continuous, bounded, with compact support, $ K_{x}(\cdot) $,  and such that
  		$ \int K_{x}(x)dx = 1  $, $  \int x K_{x}(x) dx = 0,  \int x^{2} K_{x}(x) dx > 0 $, and $ \int  K_{x}^{2}(x) dx < \infty $.  
  		\item (i) $ n b_{x} /log(n) \to \infty  $ and (ii) $ n b_{x}^{4}  \to 0.$
  	\end{enumerate}
  \end{assumption}

  For $ j = p,q $, the asymptotic linear representation of $ \widehat{UQE}_{j} $ is given as follows, 
	\begin{align}
	\widehat{UQE}&_{j}(\tau, G)  - UQE_{j}(\tau, G) - B_{j}(\tau, d, b_{y})   \nonumber  \\
	= &  \dfrac{1}{n} \sum_{i = 1}^{n} \left\{  \psi_{f_{y},j}(A_{i}; \theta_{0}, q_{\tau}, G) - \dfrac{1}{f_{Y|R}(q_{\tau}|1)}  \psi_{d,j}(A_{i}; \theta_{0}, q_{\tau}, G)  \right\} + o_{p}(n^{-1/2}b_{y}^{-1/2} + b_{y}^{2}),  \nonumber \\ 
	=&\dfrac{1}{n} \sum_{i = 1}^{n}\psi_{j}(A_{i}; \theta_{0}, q_{\tau},b_{y} )  + o_{p}(n^{-1/2}b_{y}^{-1/2} + b_{y}^{2}), \label{psi.q0}
	\end{align}
	where 
	\begin{align}
	\psi_{f_{y},j}(a; \theta_{0}, q_{\tau}, G)   := &  \dfrac{d_{j}(\theta_{0}, G) }{f_{Y|R}^{2}(q_{\tau}|1)} \dfrac{r}{Q_{0}}  \left( K_{b_{y}}(y - q_{\tau})  \right. \nonumber \\
	                                                  & \left. - \E[K_{b_{y}}(Y - q_{\tau})|R = 1]  -  \dfrac{(1(y\leq q_{\tau}) - \tau) f'_{Y|R}(q_{\tau}|1)}{f_{Y|R}(q_{\tau}|1)} \right), \label{psi.q.fy}\\
	\psi_{d,j}(a; \theta_{0}, q_{\tau}, G) :=& \left( M_{\theta,j}(\theta_{0})' \psi_{\theta}(a; \theta_{0}, q_{\tau} )  \nonumber \right.  \\ 
	& \left.  + \psi_{g,j}(a; \theta_{0}, G)  + \dfrac{(1-r)\ell(z)\Lambda_{x}(w; \beta_{0})g_{j}(x)}{1-Q_{0}}   -\dfrac{ r d_{j}(\theta_{0}, G) }{Q_{0}} \right). \label{psi.q.d} 
	\end{align}
	In the above equation, $ \psi_{\theta}(a; \theta_{0}, q_{\tau} ) $ are defined in the Supplementary Appendix, 
	\begin{align}
		M_{\theta, j}(\theta_{0})  := & \E\left[  	\dfrac{1-R}{Q_{0}}  \begin{pmatrix}
		\Lambda_{x}(W; \beta_{0}) \left( \nabla_{L,\theta_{0}}(Z) \cdot  g_{j}(X) + G_{j,\theta_{0}}(X) \right)   \\ 
		\dfrac{L_{0}(Z)}{1- L_{0}(Z)} \Lambda_{x, \beta}(W; \beta_{0}) g_{j}(X) 
	\end{pmatrix}    \right],  \label{M.theta.j} \\
	\nabla_{L, \theta}(z) :=& \begin{pmatrix} 
		\dfrac{L_{s}'(z) (1- L_{a}(z)) +L_{s}(z) L'_{a}(z)}{(1- L_{a}(z) )^{2}} \cdot k(z) \\
		\dfrac{L'_{s}(Z)}{1- L_{a}(Z)} \cdot t(Z) \\
		-\dfrac{L_{s}(Z) L_{a}'(Z)}{(1- L_{a}(Z))^{2}} \cdot t(Z)
	\end{pmatrix},
	\end{align}
	for		$ L_{j}(z)  := L(k(z)'  \gamma + t(z)'  \lambda_{j} ) $, and $ L_{j}'(z)  := L'(k(z)' \gamma  + t(z)' \lambda_{j} ) $, $ j = s,a.$ In addition, 
	\begin{align}
	    G_{q,\theta_{0}}(x)  := &  G^{'}(G^{-1}(F_{X|R}(x| 1)))^{-1} \cdot \E\left[ \dfrac{1-R}{Q_{0}} \cdot  \nabla_{L,\theta_{0}}(Z)  1(X \leq x )  \right],  \label{G_q.theta} \\
		G_{p, \theta_{0}}(x)  := &  \left.  \E\left[  \frac{1-R}{Q_{0}}   \nabla_{L,\theta_{0}}(Z)  1(X \leq x )  \right] \middle/ f_{X|R}(x|1) \right.  \nonumber \\
		&+ (G(x) - F_{X|R = 1}(x)) \left. \E\left[ \frac{1-R}{Q_{0}} \nabla_{L,\theta_{0}}(Z) I_{b_{x}} K_{b_{x}}(X - x) \right]  \middle/  f_{X|R}(x|1)^{2}, \right.  \nonumber \\
		\psi_{g,q}(a; \theta_{0}, G) :=&\E\left[ \frac{1-R}{1-Q_{0}}  \cdot  \frac{\ell(Z)  \Lambda_{x}(W; \beta_{0} )  }{G'\left(G^{-1}(F_{X|R}(X|1))\right)}  \right.  \nonumber \\ 
		&\left.\cdot  \left( \frac{(1- r)\ell(z) 1(x \leq X) }{1-Q_{0}}  - \frac{r F_{X|R}(X|1)}{Q_{0}} \right)  \right], \label{psi.x} \\
		\psi_{g,p}(a; \theta_{0}, G)  :=  & \E \left[\frac{(1-R)\ell(Z)\Lambda_{x}(W;\beta_{0})}{(1-Q_{0}) f_{X|R}(X |1) } \cdot \left( \frac{  (1-r)\ell(z)1( x \leq  X )  }{1-Q_{0}}  - F_{X|R}(X|1) \right) \right] \nonumber \\
		& +   \dfrac{(1-r )\ell(z) }{1-Q_{0}} \pi(x) - \E\left[   \dfrac{(1-R )\ell(Z) }{1-Q_{0}} \pi(X)  \right] \nonumber \\
		& - \dfrac{r - Q_{0}}{Q_{0}} \cdot  \E \left[\dfrac{(1-R)\ell(Z)\Lambda_{x}(W;\beta_{0})}{1-Q_{0}}  \cdot \dfrac{ G(X) }{f_{X|R}(X |1)}\right], \label{psi.xp} \\ 
		\pi(x)   := & \E\left[  \left. \Lambda_{x}(W;\beta_{0})  \right \vert X = x, R = 1\right] \dfrac{G(x) - F_{X|R}(x |1)}{f_{X|R}(x|1)}. \label{psi.p.pi.x}
	\end{align}

\pagebreak
\onehalfspacing{\small
	\bibliographystyle{jasa}
	\bibliography{Inoue_Li_Xu_arxiv}
}



\end{document}